\documentclass{aastex631}

\usepackage{amsmath}
\usepackage{mathrsfs}

\newcommand{\fract}[2]{\leavevmode\kern.1em
          \raise.5ex\hbox{\the\scriptfont0 #1}\kern-.1em
    \raise.15ex\hbox{\the\scriptfont0 /}\kern-.08em\lower.25ex\hbox{\the\scriptfont0 #2}}

\newcommand{\cT}{c_{\scriptscriptstyle T}}

\newcommand{\half}{{\textstyle\frac{1}{2}}}

\newcommand{\quart}{{\textstyle\frac{1}{4}}}

\newcommand{\im}{\mathop{\rm Im}\nolimits}

\newcommand{\rd}{\mathrm{d}}

\newcommand{\pderiv}[2]{\frac{\partial#1}{\partial#2}}

\newcommand{\deriv}[2]{\frac{\rd#1}{\rd#2}}

\renewcommand{\a}{{\mathbf{a}}}
\renewcommand{\b}{{\mathbf{b}}}
\newcommand{\e}{\hat{\mathbf{e}}}
\newcommand{\g}{\mathbf{g}}
\renewcommand{\k}{\mathbf{k}}

\newcommand{\m}{\mathbf{m}}

\newcommand{\boldv}{{\mathbf{v}}}

\newcommand{\bolds}{{\mathbf{s}}}
\newcommand{\x}{\mathbf{x}}
\newcommand{\A}{{\mathbf{A}}}
\newcommand{\B}{{\mathbf{B}}}
\newcommand{\C}{{\mathbf{C}}}
\newcommand{\D}{{\mathbf{D}}}
\newcommand{\E}{{\mathbf{E}}}
\newcommand{\F}{{\mathbf{F}}}
\newcommand{\I}{{\mathbf{I}}}
\newcommand{\J}{{\mathbf{J}}}
\newcommand{\N}{{\mathbf{N}}}
\newcommand{\X}{\mathbf{X}}
\newcommand{\Y}{\mathbf{Y}}
\newcommand{\M}{{\mathbf{M}}}
\renewcommand{\P}{{\mathbf{P}}}
\newcommand{\boldR}{{\mathbf{R}}}

\newcommand{\boldQ}{{\mathbf{Q}}}
\newcommand{\U}{\mathbf{U}}
\newcommand{\V}{\mathbf{V}}
\newcommand{\W}{\mathbf{W}}

\newcommand{\rhoc}{\rho_{\mathrm{c}}}
\newcommand{\rhon}{\rho_{\mathrm{n}}}
\newcommand{\cc}{c_{\mathrm{c}}}
\newcommand{\ac}{a_{\mathrm{c}}}
\newcommand{\cn}{c_{\mathrm{n}}}
\newcommand{\uc}{u_{\mathrm{c}}}

\newcommand{\wc}{w_{\mathrm{c}}}

\newcommand{\boldvc}{{\mathbf{v}_{\mathrm{c}}}}
\newcommand{\boldvn}{{\mathbf{v}_{\mathrm{n}}}}
\newcommand{\boldvs}{{\mathbf{v}_{\mathrm{s}}}}
\newcommand{\nunc}{\nu_{\mathrm{nc}}}

\newcommand{\nucn}{\nu_{\mathrm{cn}}}
\newcommand{\acn}{\alpha_{\mathrm{cn}}}
\newcommand{\RH}{\boldR_\mathrm{H}}
\newcommand{\RA}{\boldR_\mathrm{A}}
\renewcommand{\DH}{\D_\mathrm{H}}

\newcommand{\R}{\mbox{\boldmath$\mathsf{R}$}}
\renewcommand{\S}{\mbox{\boldmath$\mathsf{S}$}}
\newcommand{\Q}{\mbox{\boldmath$\mathsf{Q}$}}

\newcommand{\vdot}{{\boldsymbol{\cdot}}}
\newcommand{\vcross}{{\boldsymbol{\times}}}
\newcommand{\grad}{\mbox{\boldmath$\nabla$}}
\newcommand{\bxi}{\mbox{\boldmath$\xi$}}
\newcommand{\bkappa}{\mbox{\boldmath$\kappa$}}

\newcommand{\thth}{\hspace{1.5pt}}
\newcommand{\curl}{\grad\vcross}

\newcommand\Div{\grad\vdot\thth}

\newcommand{\kperp}{k_{\scriptscriptstyle\!\perp}}
\newcommand{\boldkperp}{\k_{\scriptscriptstyle\!\perp}}

\newcommand{\kpar}{k_{\scriptscriptstyle\parallel}}

\renewcommand{\bv}{Brunt-V\"ais\"al\"a}

\renewcommand{\leq}{\leqslant}  \renewcommand{\le}{\leqslant}
  \renewcommand{\ge}{\geqslant}

\begin{document}

\title{Wave Conversion, Decay and Heating in a Partially Ionized\\ Two-Fluid Magneto-Atmosphere}

\author[0000-0001-5794-8810]{Paul S. Cally}
\affiliation{School of Mathematics, Monash University \\
Victoria 3800, Australia}

\author[0000-0001-6373-3138]{M. M. G\'omez-M\'iguez}
\affiliation{Instituto de Astrof\'isica de Canarias\\ 38205 La Laguna, Tenerife, Spain}
\affiliation{Departamento de Astrof\'isica, Universidad de La Laguna\\ 38205 La Laguna, Tenerife, Spain}



\begin{abstract}
A ray-theoretic phase space description of linear waves in a two-fluid (charges and neutrals) magnetized plasma is used to calculate analytic decay rates and mode transmission and conversion coefficients between fast and slow waves in two dimensions due to finite ion-neutral collision frequencies at arbitrary ionization fraction. This is relevant to partially ionized astrophysical plasmas, in particular solar and stellar atmospheres. The most important parameter governing collisional effects is the ratio of the wave frequency to the neutral-charges collision frequency, $\epsilon=\omega/\nunc$, with secondary dependence on ionization fraction and wave attack angle. Comparison is made to the one-fluid magneto\-hydro\-dynamic (MHD) case, and it is found that acoustic-to-acoustic and magnetic-to-magnetic transmission through the Alfv\'en-acoustic equipartition layer is decreased by a term of $\mathcal{O}(\epsilon^2)$ relative to one-fluid (infinite collision frequency), and correspondingly acoustic-to-magnetic and magnetic-to-acoustic conversion is increased. The neutral acoustic mode is shown to dissipate rapidly as $\nunc\to\infty$. Away from the mode conversion region, dissipative decay along the remaining magneto-acoustic rays scales as $\mathcal{O}(\epsilon)$ and is found to be much more effective on magnetically dominated rays compared to acoustically dominated rays. This produces a steep jump in dissipation in mode conversion regions, where the rays change character, and can produce localized heating there and beyond. Applications to the solar chromosphere are discussed.
\end{abstract}



\section{Introduction} \label{sec:intro}
Classical magneto\-hydro\-dynamics (MHD) is a  single-fluid (1F) description of highly collisional plasmas whose electromagnetic behaviour is totally characterized by the magnetic field $\B$ \citep{GoePoe04aa}. In a partially ionized plasma, the collisions tie the neutrals to the charges and hence to $\B$. In other words, neutrals indirectly feel magnetic forces due to the collisional coupling. However, as collision frequencies reduce, at lower densities or temperatures for example, there can be some drift between species. For charge-neutral drifts, this is accounted for in 1F non-ideal MHD by the introduction of ambipolar diffusion via a generalized Ohm's law \citep{ZaqKhoRuc11aa, KhoColDia14aa}. 

Single-fluid MHD waves propagate ubiquitously in solar and stellar atmospheres spanning many density scale heights. They are believed to contribute to heating the solar chromosphere, transition region and corona \citep{McIDe-12aa,De-PolHan21rm,SriBalCal21vr}. However, the usual characterisation of MHD waves as fast, slow and Alfv\'en is no longer global in a stratified atmosphere \citep{CalGoo08aa,GooArrVan19aa,Cal22in}. Fast and slow waves may inter-convert near the Alfv\'en acoustic equipartition level where Alfv\'en and sound speeds coincide \citep{SchCal06aa}. Fast and Alfv\'en waves may also resonantly couple near the fast wave reflection height, provided the waves are not propagating in the vertical plane of the magnetic field \citep{CalHan11aa}. These conversions have implications for atmospheric heating and for interpretation of observations.

However, the solar photosphere and low chromosphere are only weakly ionized, with ionization fraction as low as $10^{-4}$ in the quiet Sun temperature minimum region, or even lower in sunspot umbrae \citep{KhoColDia14aa}. For waves whose frequency is low in comparison to the collisional frequencies, momentum and energy is efficiently exchanged between neutrals and charges (ions and electrons) and they move as a whole. Any slippage or drift between the two species is dissipative and leads to energy loss from waves. In 1F descriptions, this effect is accounted for by the introduction of ambipolar diffusion in a generalized Ohm's law, but is a natural consequence of collisional coupling between species in two-fluid (2F) models.

\citet{CalKho18aa, CalKho19aa} and \citet{KhoCal19aa} studied fast-to-Alfvén conversion in the solar chromosphere taking account of 1F ambipolar diffusion, finding that the energy of the fast mode is efficiently released before entering the conversion region. In this article, we use 2F modelling to explore the extent to which partial ionization and drift affect fast/slow mode conversion and dissipation in two dimensions (2D). The Alfv\'en wave, which is polarized in the third dimension perpendicular to both magnetic field and wavevector, is not included.

Applications are to space and laboratory plasmas. Ambipolar diffusion has been successfully applied in the modelling of neutron stars \citep{Jon87aa}, proto-planetary disks \citep{BaiSto11aa}, the interstellar medium \citep{Bra19aa} and solar and stellar chromospheres \citep{PopKep21aa}. A two-fluid model has the potential to improve the description of high frequency processes by describing the drift explicitly. This applies in particular to chromospheres \citep{PopLukKho19aa, ZhaPoeLan21aa}, the topic considered here. Dissipative effects deriving from two-fluid collisional coupling may also be of importance in solar prominences \citep{ForOliBal07aa,ForOliBal08aa} and have been shown to be more important than viscosity or thermal conductivity in the partially ionized solar atmosphere  \citep{KhoArbRuc04aa, KhoRucOli06aa}.

Waves in plasmas differ in nature depending on their frequencies relative to the  characteristic frequencies of the plasma, such as collision and gyro-frequencies.  In the lower solar atmosphere, the neutral-charges collision frequency (of around 300--$10^6$ $\rm s^{-1}$ depending on height) is typically the smallest of these natural frequencies, and we focus primarily on waves below this range. %This is discussed further in Section \ref{sec:collision}. 

In Section \ref{sec:eqns} we set out the basic equations of a two-fluid plasma, discuss the charges-neutrals and neutrals-charges collision frequencies, introduce the eikonal approximation and \citep[following][]{SolCarBal13gq} the dispersion relation, identify fast, slow and neutral-acoustic modes, and analytically show how the neutral-acoustic wavenumber disappears via complex infinity in the high collision frequency (one-fluid) limit. In Section \ref{sec:conv} a general mode conversion theory is described \citep[following][]{TraBriRic14aa} and then applied to the fast and slow two-fluid modes, obtaining analytic transmission and conversion coefficients that generalize the one-fluid formula. In Section \ref{sec:decay} an analytic weak-dissipation theory is developed that returns simple formulae for the dissipation/heating rates of the slow and and fast waves in the high and low plasma-beta regimes, and show how mode conversion is implicated in switching these on near the Alfv\'en-acoustic equipartition surface. In Section \ref{sec:conc} our results are summarized and solar implications discussed.

%%%%%%%%%%%%%%%%%%%%%%%%%%%%%%%%%%%%%%%%%%
\section{Mathematical Formulation} \label{sec:eqns}

%%%%
\subsection{Basic Equations} \label{sec:basic}
Consider a hydrogen plasma consisting of two components, the charges (ions and electrons) and the neutrals, with equilibrium densities $\rhoc$ and $\rhon$, equilibrium pressures $P_\mathrm{c}$ and $P_\mathrm{n}$, perturbed pressures $p_\mathrm{c}$ and $p_\mathrm{n}$, and fluid velocities $\boldvc$ and $\boldvn$. Then following \citet{SolCarBal13gq}, the coupled two-fluid linearized equations may be written as
\begin{subequations}\label{basiceqns}
\begin{gather}
\rhoc \pderiv{\boldvc}{t}=-\grad p_{\mathrm{c}}+\frac{1}{\mu}(\curl\b)\vcross\B-\acn(\boldvc-\boldvn),\label{mmntm c}\\[4pt]
\rhon \pderiv{\boldvn}{t}=-\grad p_{\mathrm{n}}+\acn(\boldvc-\boldvn),\label{mmntm n}\\[4pt]
    \pderiv{\b}{t}=\curl(\boldvc\vcross\B),\label{b}\\[4pt]
    \pderiv{p_{\mathrm{c}}}{t}=-\Gamma_1 P_{\mathrm{c}} \Div\boldvc,\label{p c}\\[4pt]
    \pderiv{p_{\mathrm{n}}}{t}=-\Gamma_1 P_{\mathrm{n}} \Div\boldvn, \label{p n}
\end{gather}
\end{subequations}
where $\Gamma_1=\fract{5}{3}$ is the adiabatic index, $\mu=4\pi\times10^{-7}\ \rm H\, m^{-1}$ is the magnetic permeability, 
$\acn$ is the friction coefficient, $\B$ is the background magnetic field, assumed constant or slowly varying in space, and $\b$ is the perturbed magnetic field. The 2F description is well-justified in the solar chromosphere because the electron-ion collision rate is orders of magnitude larger than the ion-neutral rate \citep{KhoColDia14aa}, making a three-fluid (3F) description unnecessary.

The 2F equations have been written down in some generality many times before, e.g., by \citet{ZaqKhoRuc11aa}, \citet{KhoColDia14aa} and \citet{BalAleCol18aa}. In particular, the pure hydrogen 2F model is described by \citet{PopLukKho19aa}. Collisions only contribute to the momentum equations in the linearized equations (\ref{basiceqns}) since the collisional terms in the full energy equations are quadratic in the velocities for an initial state where charges and neutrals are in thermal equilibrium. This causes the 2F linear waves to not conserve energy, though see the discussion in Appendix \ref{app:elas}. In the presence of a magnetic field, collisional terms produce ambipolar diffusion, included as a non-ideal term in Ohm's law in the 1F description. Collisional terms also yield other effects, such as the pressure function $\mathbf{G}$ \citep{ForOliBal07aa}. Mainly, the role of these terms in both 1F and 2F models is balancing momentum exchange and describing the drift in velocity of charges and neutrals.

There has been some discussion in the literature about whether transverse magnetic waves (Alfv\'en, slow, kink) can actually exist in the solar photosphere because of the low ionization fraction there. \citet{VraPoePan08aa} argue that they cannot, or at least not with significant amplitudes. However, this is contradicted by the results of \citet{TsaSteKop11aa}. The discrepancy was resolved by \citet{SolCarBal13aa}, who found the disagreement was largely rooted in the initial conditions applied, and that if both charges and neutrals are driven similarly, then standard MHD applies; see also \citet[Sec.~5.3.2]{BalAleCol18aa}. Recent observations of torsional Alfv\'en waves in the photosphere of a pore \citep{StaErdBoo21aa} seem to confirm this. We proceed on the assumption that Equations (\ref{basiceqns}) adequately describe linear magneto-acoustic waves in both photosphere and chromosphere, though see the further discussion in Section \ref{sec:collision}.

The only non-ideal effect included in Equations (\ref{basiceqns}) is the collisional coupling between the two species in the momentum equations alone. Other terms neglected from the equations include viscosity, radiative loss and heating, thermal conduction, Ohmic heating, ionization/recombination, the Hall effect, the battery term, and more \citep{ForOliBal07aa,ZaqKhoRuc11aa,KhoColDia14aa,PopLukKho19aa,SnoHil20vq}. 

For simplicity, we have not included gravitational terms in the perturbation equations, though gravitational stratification will be retained in background quantities such as the densities and pressures $\rhoc$, $\rhon$, $P_\mathrm{c}$ and $P_\mathrm{n}$, etc. This has the effect of excluding the {\bv} and acoustic cutoff frequencies, which are typically of the order of a few milliHertz in the low solar atmosphere, much smaller than frequencies of interest where imperfect coupling between charges and neutrals occurs. They can be added if desired \citep{SchCal06aa}, though at the expense of greater algebraic complexity that obscures the main effects. 

Similarly, other variations in the background atmosphere (temperature, pressure, magnetic field strength and direction) that are slow compared to the wavelengths of interest are retained only via position dependence of the sound and Alfvén speeds for example, and not their derivatives. Although we shall generally discuss results in the context of a gravitationally stratified stellar atmosphere, they will apply just as well to any inhomogeneous space or laboratory plasma in which the ratio of the Alfv\'en speed $a$ to sound speed $c$ varies with position, and especially if it passes through 1 on some surface.

The individual sound speeds on the charges and neutrals, $\cc$ and $\cn$, are defined by $\cc^2=\Gamma_1 P_\mathrm{c}/\rhoc$ and $\cn^2=\Gamma_1 P_\mathrm{n}/\rhon$, and the Alfv\'en speed $\ac$ of the charges alone by $\ac^2=B^2/\mu\rhoc$. It is convenient to introduce the neutral-to-charges ionization ratio $\chi=\rhon/\rhoc$, the neutral-to-total ionization fraction $\xi_n=\rho_n/\rho=\chi/(1+\chi)$, the total sound speed $c$ given by $c^2=(\cc^2+\chi\cn^2)/(1+\chi)$, and the total Alfv\'en speed $a$ defined by $a^2=B^2/\mu\rho=\ac^2/(1+\chi)$, where $\rho=\rhoc+\rho_n$. These definitions are simply algebraic conveniences, and do not represent additional assumptions or approximations. With \cite{SolCarBal13gq}, we assume a common temperature for both species, in which case $\cc^2=2\cn^2$, and $c^2=\cn^2\,(2+\chi)/(1+\chi)$. 

%%%%%%%%%%%%%
\subsection{Collision and Other Characteristic Frequencies} \label{sec:collision}
Modelling of plasmas and in particular waves in plasmas is intrinsically dependent on timescales \citep[][Sec.~II.D]{KhoColDia14aa}. The focus here is on wave frequencies below the neutral-charges collision frequency, which is the lowest of the characteristic frequencies of concern in the solar photosphere and chromosphere. At wave frequencies around or above this, one-fluid modelling may be insufficient. In this sense, collision frequencies between charges and neutrals must be understood as a scaling reference for the validity of the 1F approach, motivating the use of the more detailed 2F description as wave frequency increases.

It is conventional to introduce the charges-neutral collision frequency $\nucn=\acn/\rhoc$ and the neutral-charges collision frequency $\nunc=\acn/\rhon$. It will be seen in Sections \ref{sec:conv} and \ref{sec:decay} that $\nunc$ plays a crucial role in both mode conversion and mode dissipative decay, so it is important to gain an idea of its magnitude. 

\begin{figure}[htb]
    \centering
    \includegraphics[width=.6\textwidth]{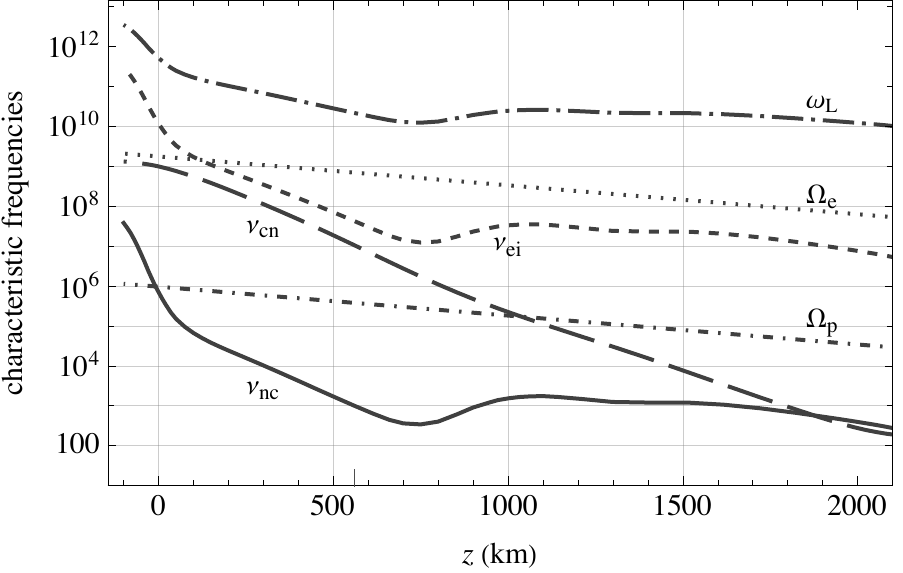}
    \caption{Approximate neutral-charges collision frequency $\nunc$ (collisions per second, full curve), charges-neutral frequency $\nucn$ (long-dashed) and electron-ion collision frequency $\nu_\text{ei}$ (short-dashed) based on the quiet atmospheric model C7 of \citet{AvrLoe08aa}. The proton ($\Omega_\text{p}$, short-chained) and electron ($\Omega_\text{e}$, dotted) gyro\-frequencies ($\rm rad\,s^{-1}$) are shown for comparison, based on the magnetic field strength $B=100 \exp[-z/600]$~G as in \citet{KhoColDia14aa}. The Langmuir frequency $\omega_\text{L}$ (long-chained) indicates the rapid timescale on which charge neutrality is imposed.  The extended tick mark at the bottom corresponds to the position of the temperature minimum.}
    \label{fig:nunc}
\end{figure}

Based on the \citet{Bra65aa} expression for the ion-neutral and electron-neutral collision coefficients \citep{PopLukKho19aa}, the total rates $\nucn=(\rho_\text{e}\nu_\text{en}+\rho_\text{i}\nu_\text{in})/\rho_\text{c}$ and $\nunc=\nucn/\chi$ may be calculated for the C7 atmospheric model of \citet{AvrLoe08aa}. This is essentially an updated version of the well-known VAL C model of \citet{VerAvrLoe81aa}. These collision frequencies are represented in Figure \ref{fig:nunc}. Translating from a chemically complex model such as C7 to a simple pure hydrogen gas necessarily requires some level of approximation; for example, we assume the ion and electron number densities are the same, $n_\text{i}=n_\text{e}$. Nevertheless, these values are comparable to the more sophisticated result of \citet{VraKrs13aa} shown in their Figure 10.

The electron-ion collision frequency $\nu_\text{ei}$, which lies over four orders of magnitude above $\nunc$, is plotted too, indicating that the ions and electrons together can indeed be modelled as a single fluid at the low frequencies addressed here. They are even more rapidly coupled by Langmuir waves of frequency $\omega_\text{L}$, which maintain charge neutrality via electron shielding, and are unrelated to collisions.

The proton and electron gyrofrequencies are also plotted in Figure \ref{fig:nunc}, based on the typical quiet Sun magnetic field strength $B=100 \exp[-z/600]$~G posited by \citet{KhoColDia14aa}, showing that the electrons are magnetized throughout. The ion gyrofrequency greatly exceeds $\nunc$ for $z\gtrsim100$ km, but is much less than $\nucn$ on $z\lesssim1000$ km. 

The situation that the electron-ion collision frequencies are lower than the electron gyro\-frequency but greater than the ion gyro\-frequency is common in dense astrophysical plasmas such as the solar photosphere and planetary ionospheres. It sees electron drift perpendicular to the magnetic field and ion drift parallel to the electric field, which is modelled within a fluid description using an anisotropic electrical conductivity tensor characterized by distinct longitudinal, Pederson and Hall\footnote{The significance of the Hall effect for solar MHD waves in one-fluid models has previously been discussed by \cite{CalKho15aa} and \citet{GonKhoCal19aa} and found to be both non-dissipative and effective only at high frequencies $\omega\gtrsim \xi_\text{i}\Omega_\text{i}$ \citep[see also][]{PanWar08aa}, where $\xi_\text{i}=\rho_\text{i}/\rho$ is the ionization fraction and $\Omega_\text{i}$ is the ion gyro\-frequency. In any case the Hall effect may be accounted for via a generalized Ohm's law in either iF or 2F models. It operates intrinsically in three dimensions (3D) as it has the property of rotating transverse wave polarizations about the magnetic field direction, which is inconsistent with the current 2D model.} terms in a generalized Ohm's law \citep{Rus03aa}.

Strong observed correlations between flows and magnetic elements such as in active region emergences \citep{CamUtzVar19aa}, Evershed flows \citep{RimMar06aa}, magnetic accumulations at granulation and supergranulation boundaries \citep{SprNorTit90aa,SchHagTit97aa}, etc., atest to the intimate coupling of plasma and magnetic field in the photosphere, despite the ions being `unmagnetized'. Indeed, at the lengths and velocities appropriate to these flows, the magnetic Reynolds number is much larger than 1, indicating an essentially `frozen-in' field.\footnote{For $B=100$ G, \citet{KubKar86aa} estimate the isotropic, Pederson and Hall conductivities as $\sigma_\parallel=19\ \rm S\,m^{-1}$ (which increases rapidly with height), $\sigma_\text{P}=1.3\ \rm S\,m^{-1}$ and $\sigma_\text{H}=4.8\ \rm S\,m^{-1}$ respectively at the quiet Sun temperature minimum. Assuming an MHD description, the magnetic Reynolds number $\mathcal{R}_m=L v_0 \mu\, \|\sigma\|\approx 2.4\times10^{-5} L v_0$ in SI units is the natural measure of how tightly tied are the magnetic field and plasma, where $L$ is a typical macroscopic length scale and $v_0$ is a typical fluid velocity. They are perfectly frozen together in the $\mathcal{R}_m\to\infty$ limit (Alfv\'en's theorem) but the field diffuses independently of the plasma flow if $\mathcal{R}_m\ll1$. The velocity and length scales characteristic of the abovementioned photospheric flows put them firmly in the $\mathcal{R}_m\gg1$ regime, explaining the observational correlations.} Similar conclusions can be drawn for sufficiently low frequency waves. Two-fluid effects become more pronounced as the wave frequency approaches $\nunc$.

\begin{figure}[tbh]
    \centering
    {\hfill\includegraphics[width=.47\textwidth]{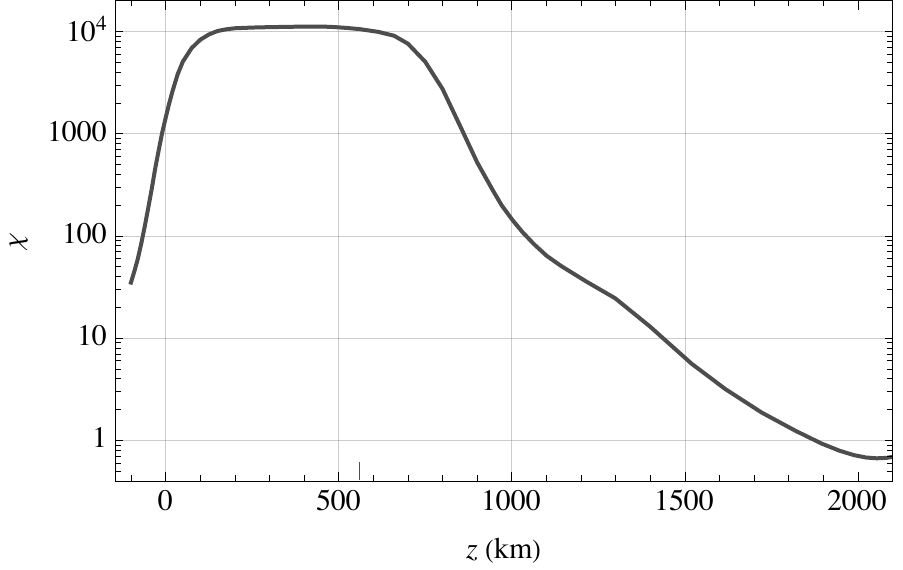}\hfill\includegraphics[width=.455\textwidth]{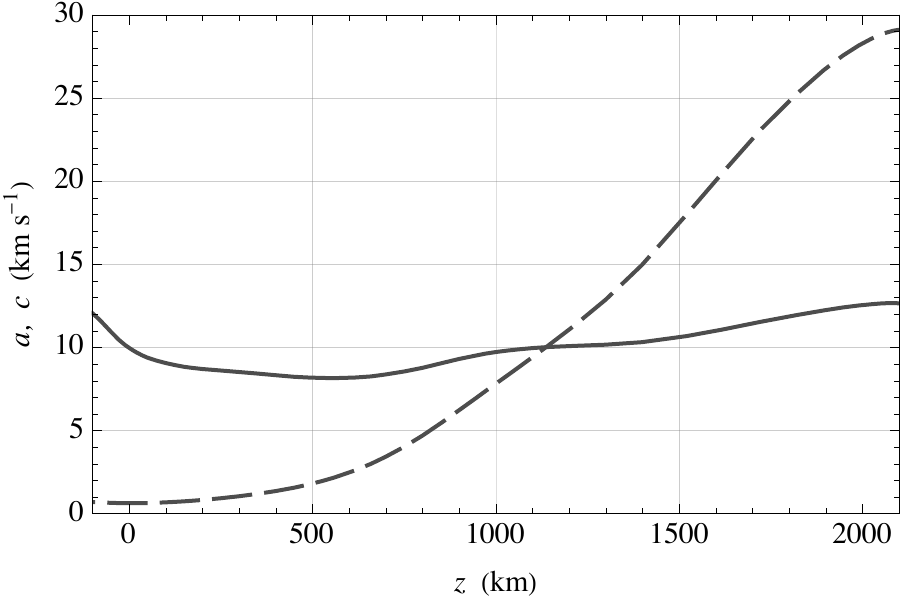}\hfill}
    \caption{Left: ionization fraction $\chi=\rhon/\rhoc$ as a function of height in model C7. Right: sound speed ($\rm km\,s^{-1}$, full curve) and Alfv\'en speed (dashed) as functions of height, assuming $B=100 \exp[-z/600]$~G again in model C7, showing the location of the equipartition level where $a=c$.}
    \label{fig:chi_ac}
\end{figure}

Representative quiet Sun ionization fraction $\chi$ (left) and sound and Alfv\'en speeds (right) in the C7 mean atmosphere are shown in Figure \ref{fig:chi_ac} as functions of height, illustrating respectively the very low degree of ionization in the photosphere and the generic behaviour that $a/c$ passes through 1 in mid-atmosphere. The Alfv\'en speed may already exceed the sound speed in intense flux elements at $z=0$, but broadly the $a=c$ level is found somewhat higher. This equipartition level is the site of fast/slow mode conversion to be discussed in Section \ref{sec:conv}. By $z=1200$ km, around where we might expect the average equipartition level to lie, the isotropic electrical conductivity is already $300\ \rm S\,m^{-1}$, so plasma/field coupling is correspondingly stronger there.

%%%%
\subsection{Eikonal Equations} \label{sec:eikonal}
Following an influential and highly cited paper by \citet{Wei62aa}, a zeroth order eikonal description is constructed assuming an $e^{i\,S}$ dependence of all perturbation variables, e.g., $\boldv=\V(\x)\, e^{i\,S} = A(\x)e^{i\,S}e^{i\,\phi}\e$, where $S(\x,t)$ is a rapidly varying phase and slow variations in the amplitudes $A$ due to background inhomogeneity are neglected as far as derivatives are concerned. Following \citet{TraBriRic14aa}, we split the vectorial amplitude $\V$ into the modulus $A(\x)$, the polarizations $\e(\x)$ which are the unit null vectors of the dispersion matrix (see Section~\ref{sec:dispmat}), and an additional phase correction $\phi(\x)$ that varies on the scales of the background. On identifying wave vector $\k=\grad S=(k_x,k_y,k_z)$ and circular frequency $\omega=-\partial S/\partial t$, this in essence sets $\grad\equiv i\,\k$ and $\partial/\partial t\equiv -i\,\omega$ when applied to the perturbations. The underlying system is assumed to be in equilibrium, so $\omega$ is strictly constant. On the other hand, the atmosphere is stratified in the vertical $z$ direction, so the $z$-component $k_z$ of $\k$ is understood to be a slowly varying function of height. Only the dominant spatial derivatives, on the scale of the wavelength, are retained.

We now specialize to the two-dimensional (2D) case of a uniform (or slowly varying) magnetic field $\B$ in the $x$-$z$ plane with $\k=(k_x,0,k_z)$ also lying in that plane. Plasma velocities in the orthogonal $y$-direction are suppressed, thereby excluding the Alfv\'en wave. Magneto-acoustic and neutral-acoustic waves remain.

The dispersion relation for the case of arbitrary collision frequencies is derived from Equations (\ref{basiceqns}) by \cite{SolCarBal13gq}, 
\begin{subequations}\label{D full}
\begin{equation}
    \mathcal{D}(\omega,\k)=D_\mathrm{i}(\omega,\k)D_\mathrm{n}(\omega,\k)+D_\mathrm{c}(\omega,\k)^2=0,  \label{D}
\end{equation}
where
\begin{gather}
    D_\mathrm{i}(\omega,\k)=\omega^3(\omega+i\,\nucn)-\omega^2 k^2(\ac^2+\cc^2)+
    \frac{\omega+i\,\nunc}{\omega+i(\nucn+\nunc)}k^4\ac^2\cc^2\cos^2\alpha,\\[4pt]
    D_\mathrm{n}(\omega,\k) = \omega(\omega+i\,\nunc)-\cn^2 k^2, \\[4pt]
    D_\mathrm{c}(\omega,\k)^2 = \frac{\omega\, \nucn\nunc}{\omega+i\,(\nucn+\nunc)}\left[\omega^3(\omega+i\,(\nucn+\nunc))-k^4\ac^2\cn^2\cos^2\alpha\right].
\end{gather}
\end{subequations}
Here $k=|\k|$ is the wave number and $\alpha$ is the attack angle between $\k$ and $\B$. 

Regarded as an expression specifying $k_z$ for given real $\omega$ and $k_x$, the dispersion relation (\ref{D}) is of sixth order, indicating that there are three modes propagating in each direction, up and down. For the most part, these can be identified as the fast and slow magneto-acoustic waves (associated with $D_\text{i}$), primarily on the charges but dragging the neutrals along via collisions, and acoustic waves ($D_\text{n}$) primarily on the neutrals. They are modified by the collision terms in their individual dispersion functions and coupled by $D_\text{c}$.

In the strongly coupled limit $\acn\to\infty$ the two species' velocities are perfectly aligned, $\boldvc=\boldvn$, and \cite{SolCarBal13gq} reduce the dispersion relation to the standard one-fluid MHD equation
\begin{equation}
    \omega^4 -(a^2+c^2)\omega^2 k^2 +a^2 c^2 k^4\cos^2\alpha=0  \label{Dmhd}
\end{equation}
in our notation (\citeauthor{SolCarBal13gq}~write it out in terms of $\rhoc$, $\cn$, $\ac$ and $\chi$). Note that the sound and Alfv\'en speeds that appear in this relation are the total versions, effectively taking account of the `mean molecular weight' $\bar\mu=(\chi+1)/(\chi+2)$ via the ionization fraction $\chi$. This is to be expected  as perfect coupling implies one fluid. 

Having introduced Equations (\ref{basiceqns}) and derived the dispersion relation (\ref{D}), \citet{SolCarBal13gq} go on to explore the different mode types for arbitrary collision frequency and various propagation directions, plotting $\omega$ against $\bar\nu$ for complex $\omega$ and real $\k$, where $\bar\nu$ is the density-weighted average of $\nunc$ and $\nucn$. This is carried out over the inter-species-collisionless to highly collisional range $0.01<\bar\nu/\cc k <100$ by solving the dispersion relation numerically for a selection of ionization ratios $\chi$. Typically, there is a bifurcation process at some $\bar\nu/\cc k=\mathcal{O}(1)$.

However, in light of the very large values of $\bar\nu$ in the solar chromosphere (see Figure \ref{fig:nunc}), and the expectation that waves of practical interest have lower frequencies than this, we depart from that course. Instead we focus on assessing the generally small-to-moderate departures from the 1F MHD modes that high collision frequencies produce. We also restrict attention to the driven case where $\omega$ is real but $k_z$ is complex. For the most part, our solutions are analytic rather than numerical, with the advantage that it is easier to discern the effects of the various parameters.

%%%%%%%%%%%%%
\subsection{Asymptotic Regimes and the Neutral Acoustic Wave} \label{sec:neutral}
Here we address the behaviour of the three wave types, especially the neutral acoustic wave, in three different regimes: (i) the case where collisions between neutrals and charges vanish, $\nucn\to0$ and $\nunc\to0$; (ii) the weak ionization limit $\chi\to0$ with $\nucn$ held fixed; and (iii) the high collision limit $\nunc\to\infty$ with ionization fraction held fixed.

%%%
\subsubsection{Low Interspecies Collision Frequency Regime}
In the absence of interspecies collisions, $\nucn\to0$ and $\nunc\to0$, the dispersion relation (\ref{D full}) decouples to give the magneto-acoustic waves on the charges, $\omega^4 -(\ac^2+\cc^2)\omega^2 k^2 +\ac^2 \cc^2 k^4\cos^2\alpha=0$, and the pure acoustic wave on the neutrals, $\omega^2=\cn^2k^2$.

%%%
\subsubsection{Low Ionization Fraction Regime}
A similar limit applies as $\chi\to\infty$, for which $\nunc\to0$ whilst $\nucn$ remains nonzero, thereby again decoupling the now-damped magneto-acoustic modes $D_\text{i}\sim\omega^2(\omega+i\,\nucn)^2 -(\ac^2+\cc^2)\omega(\omega+i\,\nucn) k^2 +\ac^2 \cc^2 k^4\cos^2\alpha=0$ from the undamped neutral acoustic mode $D_\text{n}\sim\omega^2-\cn^2k^2=0$.

Recalling that $\ac^2=B^2/\mu\rhoc=(1+\chi)a^2$ and fixing the total Alfv\'en speed $a$, the low ionization fraction regime corresponds to $\ac\gg \cc$ due to the small density of charges. To leading order then, assuming $\nunc\ll\omega\ll\nucn$, the roots of $D_\text{i}$ give
\begin{equation}\label{low_ion}
    k^2\sim \frac{\omega(\omega+i\,\nucn)}{\ac^2}\quad\text{and}\quad k^2\sim \frac{\omega(\omega+i\,\nucn)}{\cc^2}\sec^2\alpha,
\end{equation}
which are respectively an isotropic fast wave and a field-guided slow wave on the charges alone. The first vanishes in the $\chi\to\infty$ limit (i.e., $\ac\to\infty$), and both are rapidly damped, leaving only undamped acoustic waves on the neutrals. This formalizes and explains the `weakly ionized' regime examined by \citet{AlhBalFed22aa}, where magnetic field is ignored by presumption despite the model being referred to as MHD. In fact, rather than the magneto-acoustic waves not existing in this regime, they exist but vanish very quickly on a small length scale determined by the charges-neutral collision frequency.

The physical foundation of this low ionization fraction regime is that, with very few charged particles, the neutrals essentially never encounter them ($\nunc\to0$) and so are left as undamped purely acoustic waves, whilst the few charges encounter neutrals very often (large $\nucn$) and rapidly lose energy to them in collisions. This limit does not apply to `low' frequencies $\omega\lesssim\nunc$ with which we are chiefly concerned.

%%%
\subsubsection{High Collision Frequency Regime}
In the high-collision regime $\nunc\gg\omega$, the magneto-acoustic waves become the full 1F modes described by Equation (\ref{Dmhd}). However, the acoustic neutral mode does not exist in the 1F model, so what happens to it as collisions increase in 2F? 

Writing the dispersion function $\mathcal{D}$ as a monic polynomial in $k$, we have
\begin{equation}
    k^6 + c_4 k^4 + c_2 k^2 + \omega^5\,\frac{(\chi +2)^2   (\nunc(1+ \chi) -i\, \omega )^2}{2 a^2
   c^4 (\chi +1)^3 (\omega +i\, \nunc)}\sec^2\alpha  = 0,
\end{equation}
where the coefficients $c_2$ and $c_4$ need not be rendered explicitly. The final coefficient $c_0$ is shown. If $k_1^2$, $k_2^2$ and $k_3^2$ are the three roots of this bi-cubic, then\footnote{See Vieta's formulas. If $p(x)=x^n + a_{n-1}x^{n-1} + \ldots + a_0$ is a monic polynomial of degree $n$, with roots $x_1$, \ldots, $x_n$, then %the polynomial may be expressed as a product of linear factors 
$p(x) = \prod_{i=1}^n(x-x_i)$. Expanding brackets to find the term independent of $x$ reveals that the product of the roots $x_1 x_2 \ldots x_n = (-1)^n a_0$.}
\begin{equation}
    k_1^2\,k_2^2\,k_3^2=-c_0 \sim i\,\frac{\omega^5(\chi+2)^2 \sec^2\alpha}{2a^2 c^4(\chi+1)}\,\nunc
\end{equation}
for $\nunc\gg\omega$ with other variables including $\chi$ held fixed. But we know that $k_1^2$ and $k_2^2$ are real and finite in this limit, as specified by Equation (\ref{Dmhd}). In fact, by similar reasoning from Equation (\ref{Dmhd}), $k_1^2 k_2^2\sim (\omega^4/(a^2c^2))\sec^2\alpha$. Hence, $k_3^2 \sim i\,\omega\,\nunc(\chi+2)^2/(2c^2(\chi+1))$ for the remaining (neutral acoustic) mode, which is asymptotically pure positive-imaginary, and so
\begin{equation}\label{k3}
    k_3\sim\pm(1+i)\,\omega^{1/2}\,\nunc^{1/2}\frac{\chi+2}{2c\,(\chi+1)^{1/2}}
    = \pm\frac{1+i}{2\epsilon^{1/2}}(\chi+2)^{1/2} \frac{\omega}{\cn}
\end{equation}
as $\nunc\to\infty$, where $\epsilon=\omega/\nunc$ is typically small in the low atmosphere for low (mHz) and even high (Hz or tens of Hz) frequencies (see Figure \ref{fig:nunc}). This is very different from the free acoustic neutral wave, for which $k=\omega/\cn$. Equation (\ref{k3}) explains how the acoustic neutral waves disappear to complex infinity along $\arg k_3=\pi/4$ and $-3\pi/4$ in the strong coupling limit, leaving only the magneto-acoustic modes. In essence, the wave is simultaneously slowed and damped by the collisions.

%%%%
\subsection{Dispersion Matrix} \label{sec:dispmat}

It is convenient to introduce the \emph{drift velocity} of the neutrals with respect to the charges, $\epsilon\, \boldvs=\boldvn-\boldvc$, with $\epsilon=\omega/\nunc$ as above. The dimensionless factor $\epsilon$ appears naturally in the analysis, and serves here as a scaling of the drift. No assumptions about size of $\epsilon$ are made at this stage. 

With the eikonal ansatz applied to the basic equations (\ref{basiceqns}), it is a simple matter to eliminate all perturbation variables, including $\boldvs$, in favour of the velocity of the charged fluid $\boldvc=(\uc,\wc)^T$, leaving the $2\times2$ matrix equation $\mathbf{R}\boldvc=\mathbf{0}$. In the process, it is found that
$\boldvs = i\,\omega^{-2}\chi^{-1}\left((\cc^2\,\k\,\k+\ac^2\,\boldkperp \boldkperp)\,\vdot\,\boldvc-\omega^2\boldvc\right)$,
where $\boldkperp$ is the component of $\k$ perpendicular to the magnetic field.
The determinant of $\mathbf{R}$ yields the dispersion function $\mathcal{D}$. The coefficients of matrix $\mathbf{R}$ are set out in Appendix \ref{app:R}.

It is important to note that $\mathbf{R}$ is non-Hermitian, leading to dissipative behaviour. There are two common ways to proceed from here.

%%%%
\subsection{Approach I: Complex Wave Vector}
In the strong coupling limit $\epsilon\to0$, the dispersion matrix $\boldR$ is real symmetric, and $k^2$ therefore real. This is also apparent from the one-fluid MHD dispersion relation (\ref{Dmhd}). Hence, $k_z$ may be either real (travelling wave) or pure imaginary (evanescent).

\begin{figure}
    \centering
    \includegraphics[width=\textwidth]{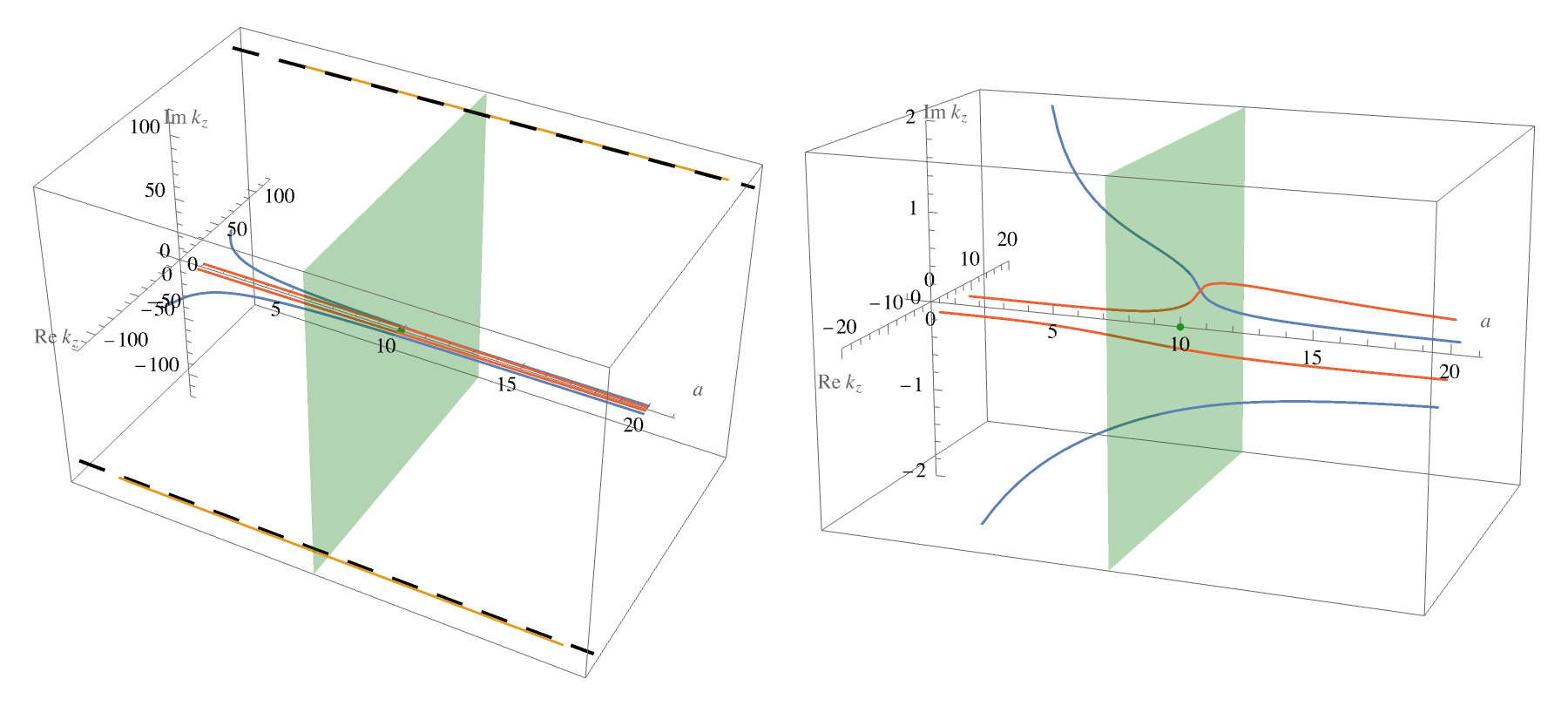}
    \caption{Complex vertical wave number $k_z$ ($\rm Mm^{-1}$) (fast: red curves; slow: blue; neutral: orange) as a function of total Alfv\'en speed $a$ ($\rm km\,s^{-1}$) for a 6 mHz wave with $k_x=1\,\rm Mm^{-1}$, total sound speed $c=10\,\rm km\,s^{-1}$, magnetic field inclination from the vertical $\theta=25^\circ$, neutral-to-charges density ratio $\chi=\rhon/\rhoc=10^3$, and frequency ratio parameter $\epsilon=\omega/\nunc=0.3$. Left: a view showing all six modes, including the highly dissipative neutral acoustic modes, for which the black dashed lines represent the asymptotic formula Eq.~(\ref{k3}), which is seen to be highly accurate. The green plane and green dot indicate the location of the equipartition level $a=c$. Right: a zoomed rendering of the same case, with the $\im k_z$ axis stretched by an order of magnitude to better display the close interaction and mutual effects of the fast and slow magneto-acoustic modes near $a=c$. } \label{fig:kz3D}
\end{figure}

This is no longer true for non-zero $\epsilon$. The non-Hermitian nature of the dispersion matrix yields complex $k_z$, which can be interpreted as representing spatial decay due to dissipation and energy loss, in addition to any evanescent behaviour. For example, the six dispersion curves derived from Equation (\ref{D}) represented in Figure \ref{fig:kz3D} for a particular choice of parameters set out in the caption clearly exhibit complex $k_z$. The real and imaginary parts of $k_z$ are plotted against the total Alfv\'en speed $a$ with fixed $\omega$, $k_x$, sound speed $c$, ionization fraction $\chi$ and frequency ratio parameter $\epsilon=\omega/\nunc=0.3$. The two inner modes are the fast waves, the intermediate ones are the slow wave, and the outermost ones are the acoustic waves on the neutrals, which are located just as predicted by Equation (\ref{k3}). 

The neutral waves are indeed extremely dissipative, and would decay over very short distances in most circumstances. The fast and slow waves on the other hand exhibit  small imaginary parts for the case shown, which decrease as $\epsilon$ decreases.

%%%%
\subsection{Approach II: Real Rays and Weak Dissipation}
A second approach is to adapt classical ray theory, where $k_z$ is real along rays. This is the weak dissipation approach set out in Sec.~3.5.1 of \citet{TraBriRic14aa}. The rays themselves are derived from the Hermitian part of $\boldR$, i.e.,  $\RH=\half(\boldR+\boldR^\dagger)$. Conversely, the dissipation is associated with the small skew-Hermitian part, $\RA=\half(\boldR-\boldR^\dagger)$, which is proportional to $\epsilon$. The dagger indicates the conjugate transpose. This approach is clearly valid only for the fast and slow waves, and not the highly dissipative acoustic neutral waves which shall henceforth be ignored.

In classical non-dissipative ray theory, the wave field is approximately reconstructed using geometrical optics, where rays are propagated throughout space and amplitudes calculated based on focusing and defocusing of these rays \citep{Wei62aa}. This breaks down at caustics, where rays cross, though there are techniques for handling these \citep[e.g., Chap.~5 of][]{TraBriRic14aa,LopDod22aa}. Such solutions do not extend beyond turning points to evanescent regions, though a complex ray theory \citep{ChaLawOck99aa} can potentially access them. 

Nevertheless, the aim here is not wave field reconstruction, but rather to answer two questions:
\begin{enumerate}
    \item How much energy is exchanged in the near-collision of the two wave types in phase space near $a=c$?
    \item To what extent are fast and slow rays diminished (beyond any geometric focusing or defocusing) by collisions between the charged and neutral fluids?
\end{enumerate}
The answer to the first question depends solely on $\RH$, using a method explained concisely by \cite{TraKauBri03aa} and at greater length by \cite{TraBriRic14aa}, Chap.~6. Answering the second question requires both $\RH$ and $\RA$ and a result from weak dissipation theory \citep[Sec.~3.5.1]{TraBriRic14aa}.

%%%%%%%%%%%%%%%
%\pagebreak[4]
\section{Mode Transmission and Conversion}\label{sec:conv}
\subsection{Ray-Based Method for Local Mode Transmission and Conversion}\label{sec:TKB}
Although standard eikonal methods break down in the neighbourhood of mode conversion regions, the ray geometry in those regions can be employed to derive a local wave description that matches between the incoming and outgoing waves. This is required because, near to the conversion point, the polarization of the rays changes rapidly \citep{TraKauBri03aa}. In its most fundamental form, the analysis rests on an Hermitian dispersion matrix of the so-called `normal' form
\begin{equation}\label{Dtkb}
    \D=
    \begin{pmatrix}
        D_a & \tilde\eta \\
        \tilde\eta^* & D_b
    \end{pmatrix},
\end{equation}
where the diagonal elements are real and the superscripted star denotes the complex conjugate. It is assumed that the coupling coefficient $\tilde\eta$ is negligible compared to the diagonal entries, and hence that the dispersion function $\mathcal{D}=\det\D\approx D_a D_b$, except in the conversion region. This shows that the dispersion relation $\mathcal{D}=0$ reduces to $D_a=0$ or $D_b=0$, which are therefore the individual dispersion relations for the two distinct modes where they are uncoupled. 

\begin{figure}
    \centering
    \includegraphics[width=\textwidth]{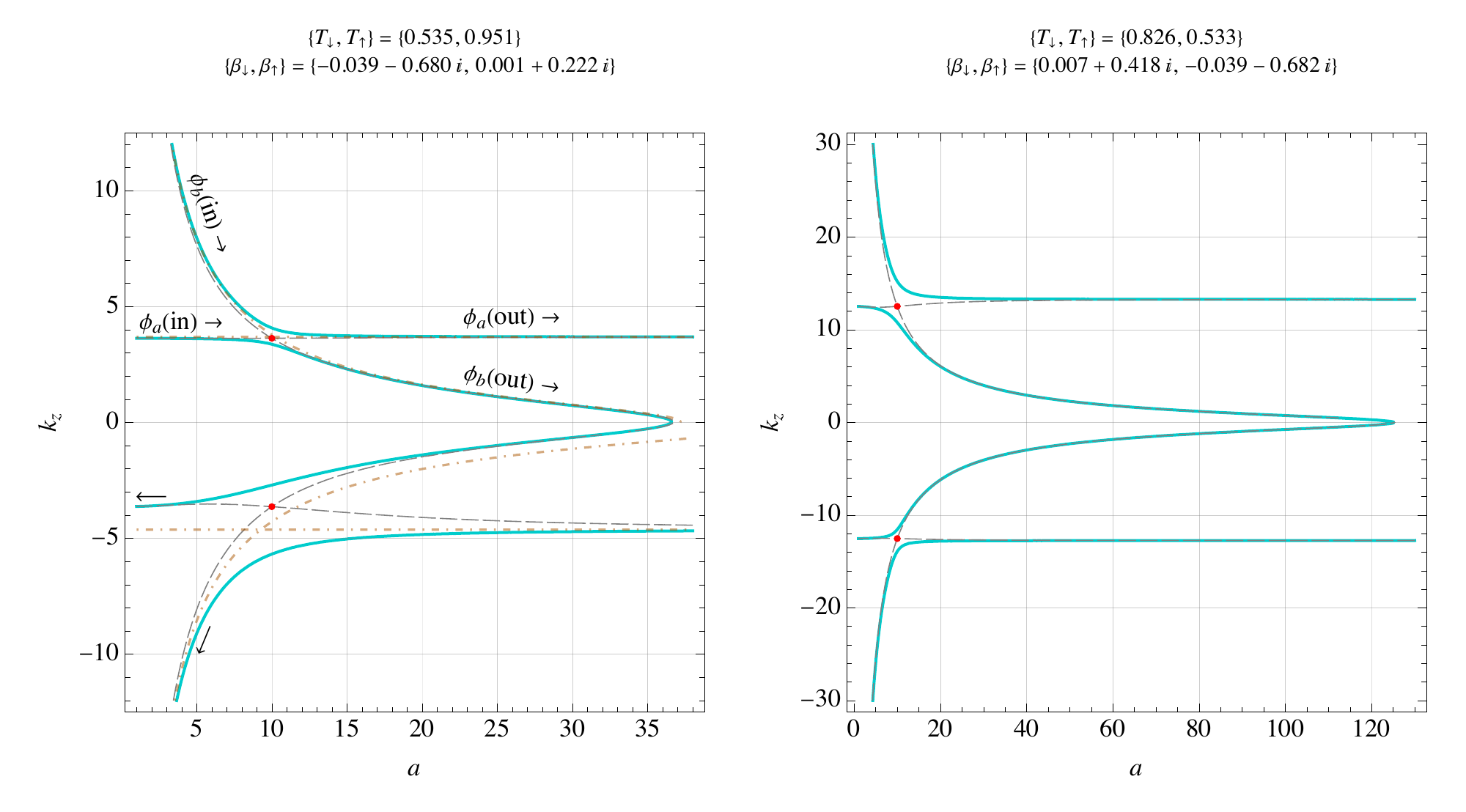}
    \caption{
    Left: a representative image of the dispersion curves in Alfv\'en speed-vertical wave number space $(a,k_z)$ ($\rm km\,s^{-1}$ and $\rm Mm^{-1}$ respectively) for the strong collision limit $\epsilon=0$. Specifically, it corresponds to a 6 mHz wave with $k_x=1\,\rm Mm^{-1}$, $c=10\,\rm km\, s^{-1}$, $\theta=25^\circ$ and $\epsilon=0$. The cyan curves are the full dispersion curves $\mathcal{D}=0$, and the black dashed curves are $D_a=0$ (the acoustic branch) and $D_b=0$ (magnetic). These cross at the star points (red dots) at $a=c$, $k_z=\pm\sqrt{\omega^2/c^2-k_x^2}$. The incoming and outgoing acoustic  and magnetic branches are labelled around the upper star point. The energy transmission coefficient at each star point ($T_\downarrow$ for the point at negative $k_z$ and $T_\uparrow$ for $k_z>0$) as well as the conversion coefficient $\beta$ for the case $h=150$ km are written above the frame; see Equations~(\ref{T}) and (\ref{tau_beta}) and the discussion in Section \ref{sec:2F}. For comparison, the $D_a=0$ and $D_b=0$ loci deriving from the saddle point placement of the star points (Equation~(\ref{saddle})) are shown as orange chained curves. Right: same, but for a 20 mHz wave and $\theta=-15^\circ$. The saddle point loci are not shown.
    }
    \label{fig:star}
\end{figure}

Typically, the conversion is associated with an avoided crossing of the (full) dispersion curves in phase space, but $D_a$ and $D_b$ instead cross, representing the alternate connectivity to that implied by $\mathcal{D}=0$. This is illustrated in Figure \ref{fig:star}. The important crossing point is referred to as the `star point', and it is where the transmission and conversion coefficients are calculated.

Under these conditions, the general theory requires 
that first $\tilde\eta$ should be scaled using the physical/phase Poisson bracket (the Jacobian of the transformation between $(z,k_z)$ and $(D_a,D_b)$ spaces)
\begin{equation}\label{poisson}
    \mathcal{B}=\{D_a,D_b\} = \pderiv{D_a}{z}\pderiv{D_b}{k_z}-\pderiv{D_a}{k_z}\pderiv{D_b}{z},
\end{equation}
where we have taken the liberty of anticipating that $z$ and $k_z$ are the physical and phase space coordinates of interest. Specifically, we define
\begin{equation}\label{eta}
    \eta = \frac{\tilde\eta}{|\mathcal{B}|^{1/2}}.
\end{equation}

Then the \emph{transmission coefficient} $\tau$ of either ray following $D_a=0$ or $D_b=0$ through the star point and the \emph{conversion coefficient} $\beta$ taking the avoided crossing route are\footnote{There is a sign convention difference in $\beta$ between \cite{TraKauBri03aa} and \cite{TraBriRic14aa}. We adopt the latter.}
\begin{equation}\label{tau_beta}
    \tau = e^{-\pi |\eta|^2}, \qquad \beta = -\frac{(2\pi\,\tau)^{1/2}}{\eta\,\Gamma(-i\,|\eta|^2)}.
\end{equation}
All coefficients are evaluated at the star point $(z,k_z)_*$ where $D_a=D_b$.
The corresponding energy transmission and conversion coefficients are $T=\tau^2$ and $C=|\beta|^2$. Using the gamma-function identity $|\Gamma(i\,y)|^2=\pi/(y\sinh(\pi y))$ for real $y$ \citep[property 6.1.29]{AbrSte65aa} 
immediately confirms that $T+C=1$, thereby conserving energy.

To understand the meanings of $\tau$ and $\beta$, consider the case illustrated in Figure \ref{fig:star}. Representing the wave solutions by $\phi_a$ and $\phi_b$, the outgoing waves after passing through the conversion region are
\begin{subequations}\label{in out}
\begin{align}
    \phi_\mathrm{ac}(\text{out}) &= \tau\, \phi_\mathrm{ac}(\text{in}) - \beta^* \phi_\mathrm{mag}(\text{in}), \\[4pt]
    \phi_\mathrm{mag}(\text{out}) &= \tau\, \phi_\mathrm{mag}(\text{in}) + \beta \,\phi_\mathrm{ac}(\text{in}).
\end{align}
\end{subequations}
The convention is that the converted branch turning right (acoustic to magnetic in this instance) inherits the conversion factor $\beta$, whilst that turning left (magnetic to acoustic) is assigned $-\beta^*$.

Clearly, $0<\tau\leq 1$, representing a partial transmission, but $\beta$ is in general complex, indicating both a change in mode amplitude and phase of the converted rays. Knowledge of the phase of the converted ray is essential in reconstructing the wave field, and in particular recovers information about interference.

%%%%
\subsection{Application to the Two-Fluid Equations}\label{sec:2F}
Mode conversion is based on the idea that two rays become close in a small region in phase space, where they resonantly interact. Far from this region, they are practically independent. Figure \ref{fig:star} shows visually that the rays have an avoided crossing in a small region in the neighbourhood of the star point, where mode conversion is expected. By assuming weak dissipation, ray propagation is enclosed in the Hermitian part of $\boldR$, 
\begin{equation}\label{RH}
    \RH=
    \begin{pmatrix}
        \omega ^2-a^2 k_z^2 -c^2 k_x^2 & (a^2-c^2) k_xk_z
   + i\,\epsilon\,E \\
 (a^2-c^2) k_xk_z
    -i\,\epsilon\,E & \omega ^2-a^2 k_x^2  -c^2 k_z^2 
    \end{pmatrix}
    =\boldR_0+i\,\epsilon\,E \,\J,
\end{equation}
where $\boldR_0$ is real symmetric, $\J=\bigl(\begin{smallmatrix}0&1\\-1&0\end{smallmatrix}\bigr)$ and
\begin{equation}\label{E}
    E = \frac{a^2 c^2 k^2 (\chi +1)  \left[\left(k_x^2-k_z^2\right)\sin 2
   \theta  +2 k_x k_z\cos 2 \theta  \right]}{4 (\chi +2) \omega ^2}.
\end{equation}
This is not in the required form of Equation (\ref{Dtkb}), since the diagonal elements do not represent the distinct modes. It is first necessary to determine the asymptotic decoupled modes, and to rotate the matrix to place their individual dispersion functions on the diagonal. This naturally yields $\tilde\eta$ in the superdiagonal entry.

It is therefore necessary to find -- or rather select -- the star points, associated `decoupled' dispersion functions $D_a$ and $D_b$, and the off-diagonal coupling term $\tilde\eta$. The role of the star point in mode conversion is as the centre of a first order Taylor expansion that permits a local wave analysis to be performed allowing asymptotic matching between the incoming and outgoing eikonal waves. This will be more or less accurate depending on the width of the avoided crossing gap. The precise position of the star point can be chosen in several ways, none of which seem \emph{a priori} superior to the others. 

An obvious choice is to set the star point to be the saddle point of the full dispersion function $\mathcal{D}=D_a D_b-|\tilde\eta|^2$, which should be close to the saddle point of $D_aD_b$. This is routinely used by \cite{TraBriRic14aa}, and in our case results in the two star points 
\begin{equation}\label{saddle}
(a,k_z)_{*\text{sad}} =
\left(c\,\sqrt{\frac{\omega(\omega\pm c\, k_x\sin\theta)}{\omega^2+c^2 k_x^2\pm2c\, k_x\omega\sin\theta}},\,\,
\mp\frac{\omega}{c}\sec\theta-k_x\tan\theta\right).
\end{equation}
The $D_a$ and $D_b$ uncoupled dispersion functions are chosen to be the separatrices passing through the saddle point.

However, we make a different choice that is algebraically simpler, and which also results in the curves $D_a=0$ and $D_b=0$ better matching to $\mathcal{D}=0$ away from the conversion region; see the orange chained loci for the saddle separatrices in Figure ~\ref{fig:star}, which clearly do not match well. Again, it is important to understand the precise selection will make very little difference for a narrow avoided crossing gap, but will result in slightly different transmission and conversion coefficients for wider gaps, for which in any case the linear Taylor expansion is less accurate.

First, the eigenvectors of $\RH$ are calculated. In the limit $a\ll c$, these are $(k_x,k_z)$ for the acoustic wave and $(k_z, -k_x)$ for the magnetic wave. Similarly, for $a\gg c$ the eigenvectors are $(\sin\theta,\cos\theta)$ for the acoustic case and $(\cos\theta,-\sin\theta)$ for the magnetic case. These are conveniently joined by an \emph{ad hoc} ramping between the two regimes using $a^2$ and $c^2$. If $\k=k\,(\sin\psi,\cos\psi)$, we define
\begin{equation}\label{Y}
    \Y=
   N^{-1} \begin{pmatrix}
        a^2 k \sin \theta_m +c^2 k_x & a^2 k \cos \theta_m +c^2 k_z \\
 a^2 k \cos \theta_m +c^2 k_z & -a^2 k \sin \theta_m -c^2 k_x
    \end{pmatrix}
\end{equation}
where $\theta_m=\theta$ if $\kpar=k_x\sin\theta+k_z\cos\theta\ge0$, and $\theta+\pi$ otherwise. The choice of $\theta_m$ is so as to rotate the eigenvectors as little as possible for the interpolation, keeping the attack angle $\alpha=\psi-\theta_m$ in $[-\pi/2,\,\pi/2]\!\!\!\mod2\pi$. The change in polarization across the conversion region is clearly small if $\alpha$ is small, also enhancing the accuracy of the method. Henceforth, we drop the subscript `m' on $\theta$; it is to be understood that the sense of the magnetic field direction $\theta$ (which has no physical implications) is always chosen so that $\kpar\ge0$.

The normalization factor
$N=[(c^2k_z + a^2k\cos\theta)^2+(c^2k_x+ a^2k\sin\theta)^2)]^{1/2}$ makes $\Y$ unitary, $\Y \Y^T=\mathbf{I}$. 

%A further consideration regarding the disjoint dispersion functions $D_a$ and $D_b$ 

Defining $\V=\Y\boldv$ then transforms $\RH\boldv=\mathbf{0}$ to $\DH\V=\mathbf{0}$, where $\DH=2N^2\Y\RH\Y^T$ (a convenient scaling). The matrix $\DH$ is now of the required form set out in Equation (\ref{Dtkb}). Specifically, with $k^2=k_x^2+k_z^2$,
\begin{multline}\label{Da}
    D_a =
    4 a^2 c^2 k \left(\omega ^2-c^2 k^2\right) \left(k_x\sin \theta 
   +k_z\cos \theta  \right)+a^2 c^2 (a^2-c^2)  \cos 2 \theta
    \left(k_x^4-k_z^4\right)-c^2 k^4 \left(a^4+a^2 c^2+2
   c^4\right)\\
   +k^2 \left(2 \omega ^2 \left(a^4+c^4\right)+2 a^2 c^2 k_x k_z
   \left(c^2-a^2\right) \sin 2 \theta  \right),
\end{multline}
\begin{multline}\label{Db}
    D_b =
4 a^2 c^2 k \left(\omega ^2-a^2 k^2\right) \left(k_x\sin \theta 
   +k_z\cos \theta  \right)-a^2 c^2 (a^2-c^2)  \cos 2 \theta
    \left(k_x^4-k_z^4\right)-a^2 k^4 \left(2 a^4+a^2
   c^2+c^4\right)\\
   +k^2 \left(2 \omega ^2 \left(a^4+c^4\right)+2 a^2
   c^2k_x k_z (a^2-c^2)  \sin 2 \theta  \right),
\end{multline}
exhibiting a nice symmetry, and
\begin{multline}\label{tilde eta}
    \tilde\eta =
    -\frac{2 a^2 c^2 k^3 \left(k_x\cos \theta  -k_z\sin \theta 
   \right) (a^2 +c^2)\left( k_x \sin \theta  + k_z
    \cos \theta  + k\right)
   \left(k \left(a^4+c^4\right)+2 a^2 c^2 \left(k_x\sin \theta 
   +k_z\cos \theta  \right)\right)}{a^4 k^2+c^2 \left(2 a^2 k k_x
   \sin \theta  +2 a^2 k k_z \cos \theta  +c^2
   k^2\right)}\\
   -i\,\epsilon\,\frac{  a^2 c^2 k^3 (\chi +1)  \left(\sin
   2 \theta  \left(k_x^2-k_z^2\right)+2k_x
   k_z \cos 2 \theta  \right) \left(k \left(a^4+c^4\right)+2 a^2 c^2 \left(k_x\sin
   \theta  +k_z\cos \theta  \right)\right)}{2 (\chi +2)
   \omega ^2}.
\end{multline}

Conveniently, $D_a$ and $D_b$ are independent of $\epsilon$ and simultaneously vanish precisely at the equipartition point $a=c$. Specifically,
\begin{equation}\label{star}
(a,k_z)_* =
\left(c,\,\,\pm\sqrt{\frac{\omega^2}{c^2}-k_x^2}\,
\right),
\end{equation}
which is typically close to but algebraically simpler than the saddle points.
Two representative cases are illustrated in Figure \ref{fig:star}, showing the avoided crossings and the alternate connections of the $D_a$ and $D_b$ dispersion curves, representing the acoustic and magnetic behaviours respectively.

Evaluating the Poisson bracket at the star point, which quantifies the angle between the $D_{a,b}$ loci, we find
\begin{equation}\label{Bpoiss}
    \mathcal{B}_* = 128\, h^{-1} c^{12} k^6 k_z (2-\cos \alpha)\cos^4\left(\alpha /2\right)  ,
\end{equation}
where %$\alpha=\psi-\theta\in[-\pi/2,\pi/2]\!\!\!\mod2\pi$ is the attack angle between the wave vector $\k=k\,(\sin\psi,\cos\psi)$ and magnetic field $\B$ at that point and
$h=(d\ln(a^2/c^2)/dz)^{-1}$ is the scale height of the conversion layer. If $B$ is uniform, $h$ is just the pressure scale height. We interpret $h$ as the distance in $z$ over which the resonant coupling of wave modes occurs. 

It is then straightforward to calculate
\begin{equation}\label{eta_star}
    \eta_* = -\frac{k \sqrt{| h| }\sin \alpha  }{\sqrt{2} \sqrt{2-\cos \alpha } \sqrt{\left| k_z\right| }}\left(1+\half\, i\,\epsilon\, \bar\mu
     \cos  \alpha
  \right),
\end{equation}
where $\bar\mu=(\chi+1)/(\chi+2)$ is the mean particle mass in units of the hydrogen atom mass $m_\text{H}$. The wave amplitude transmission coefficient $\tau=\exp(-\pi|\eta_*|^2)$ is therefore affected only at $\mathcal{O}(\epsilon^2)$ by collisions. 

The energy transmission coefficient is
\begin{equation}\label{T}
    T =\tau^2= \exp\left[
    -\pi\, k \left|h_s\right| \frac{\sin^2\alpha}{2-\cos\alpha}\left(
    1+\quart\,\epsilon^2\,{\bar\mu}^2\cos^2\alpha
    \right)
    \right]_*,
\end{equation}
where $h_s=h\sec\psi$ and $\psi$ is the angle the wave vector $\k$ makes to the vertical at the star point. That is, $h_s$ is the distance traversed by the oblique ray in crossing the horizontal slab of thickness $h$. The energy conversion coefficient is most conveniently given by $C=1-T$.

In addition to energy transmission and conversion, the converted rays also pick up a phase change via the complex conversion coefficient $\beta = -(2\pi\,\tau_*)^{1/2}/(\eta_*\,\Gamma(-i\,|\eta_*|^2))$, as per Equations (\ref{in out}), which is affected at $\mathcal{O}(\epsilon)$

The dependence of transmission and conversion on $\epsilon$ is new. The factor $\bar\mu\in[\half,1]$ is a relatively weak function of ionization fraction, although the exponential in $\tau$ amplifies it. The $\cos^2\alpha$ in $T$ term favours small attack angle, but the overall angular frictional factor is $\sin^2\alpha\cos^2\alpha/(2-\cos\alpha)$, which is maximal at about $41^\circ$ (see Figure \ref{fig:eta_diff}). However, the most important effect is via the squared frequency ratio $\epsilon^2=\omega^2/\nunc^2$. When this reaches $\mathcal{O}(1)$, transmission is significantly reduced and conversion enhanced by two-fluid frictional effects compared to the one-fluid result.

%%%%
\subsection{One-Fluid Mode Transmission and Conversion}\label{sec:1F}

\begin{figure}
    \centering
    \includegraphics{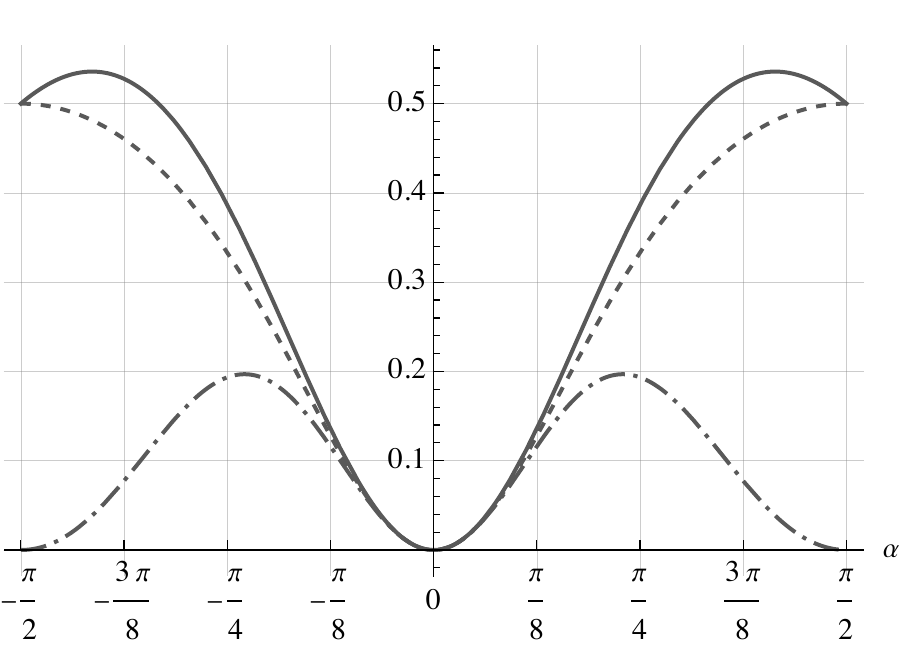}
    \caption{Plots of the angular factor $\sin^2\alpha/(2-\cos\alpha)$ in $T$ (full curve) and that of \citet{SchCal06aa}, $\sin^2\alpha/(1+\sin^2\alpha)$ (dashed), for attack angle $-\frac{\pi}{2}<\alpha<\frac{\pi}{2}$. The chained curve depicts the overall frictional angular factor $\sin^2\alpha\cos^2\alpha/(2-\cos\alpha)$.}
    \label{fig:eta_diff}
\end{figure}

Mode transmission in a similar one-fluid magneto-atmosphere was addressed using a related technique by \cite{Cal06aa} and \cite{SchCal06aa}, who obtained a similar expression for $T$  \cite[Eq.~(26), first line]{SchCal06aa} that
differs only slightly from that found here for $\epsilon=0$. Specifically, where we have the attack angle dependence of $\sin^2\alpha/(2-\cos\alpha)$, they had $\sin^2\alpha/(1+\sin^2\alpha)$. The difference between the two is illustrated in Figure \ref{fig:eta_diff}. It arises from the slightly different way the two asymptotic regimes $a\ll c$ and $a\gg c$ are bridged. The difference is entirely negligible for small $\alpha$, and quite limited elsewhere. \cite{SchCal06aa} did not calculate the phase boost in MHD, as only $|\eta_*|$ was found with their method, not $\eta_*$.

The energy transmission coefficient $T$ and the conversion coefficient $\beta$ are given at the top of Figure \ref{fig:star} (left panel) for a 6 mHz wave with $k_x=1\,\rm Mm^{-1}$, $c=10\,\rm km\, s^{-1}$, $\theta=25^\circ$, $\epsilon=0$ and $h=150$ km (typical of the solar chromosphere). The coefficients illustrate that transmission is enhanced when the avoided crossing gap is narrow, and that there can be a substantial jump in phase in the converted rays. Figure \ref{fig:star} also illustrates the typical MHD fast wave reflection process that occurs where $k_z=0$. 

The right panel of Figure \ref{fig:star} also shows that narrow (large $T$) and wide (small $T$) avoided crossings swap places for negative $\theta$, and that higher frequency fast waves reach higher in the atmosphere.

%%%%%%%%%%%%%%%
\section{Mode Decay along Rays and Local Heating}\label{sec:decay}
We now turn to calculating the rate of dissipation along rays due to the collision terms. It is of interest to see how this differs for the different ray types and on location, specifically on $a/c$. For comparison with simulations, any contrast between $a<c$ and $a>c$ will be important.

The relative rate of decay of the amplitude of a ray as it traverses a region with non-zero dissipation $\epsilon$ is described by the imaginary part of the eikonal $S=S_r+i\,S_i$:
\begin{equation}\label{decay}
    \gamma=-\deriv{\x}{t}\,\vdot\,\grad S_i=-\k_i\,\vdot\,\deriv{\x}{t},
\end{equation}
(i.e., there is an $e^{\gamma t}$ multiplicative time dependence factor, with $\gamma<0$), where $d\x/dt$ is the ray propagation velocity in physical space, also known as the group velocity. This can be calculated by direct solution of the full dispersion relation (\ref{D full}) to find $\k$ and application of the ray equations, or calculation of the group velocity, to find $d\x/dt$. Here, we implicitly assume that the rays themselves are real, deriving from the Hermitian part of the dispersion matrix, so this is in effect a perturbation result. 

This is made more formal in Appendix \ref{app:weakz} for the weak dissipation regime. It is found that
\begin{equation}\label{gamma}
  \gamma_\lambda = - \epsilon\,\frac{\e_\lambda^\dagger\,\vdot\,\boldQ(\x,\k_r)\,\vdot\, \e_\lambda}{\partial D_\lambda/\partial\omega} 
\end{equation}
to $\mathcal{O}(\epsilon)$, where $D_\lambda=\e_\lambda^\dagger\,\vdot\,\boldR_0\,\vdot\,\e_\lambda$ is the specific dispersion relation branch $\lambda$ of the ray in question, $\boldR_0$ is given by Equation (\ref{RH}) with $\epsilon=0$, $\e_\lambda$ is the unit null vector of $\boldR_0$ on that ray branch, and $\boldQ=-i\,\epsilon^{-1}\RA$ is a positive-definite real symmetric $\mathcal{O}(\epsilon^0)$ matrix. The subscript $\lambda$ ranges over $\alpha$, $\beta$, \ldots, corresponding to the different roots $\k$ of the dispersion function, for example the cyan curves in Figure \ref{fig:star}. This has the advantage that $\k_i$ need not be explicitly calculated (it can only be found numerically) and that only the much simpler MHD dispersion relation (\ref{Dmhd}) is required, apart from the straightforward appearance of an explicit expression for $\boldQ$. 

Of course the null vectors only have meaning on the dispersion manifold. If $\boldR_0(\x,\k)=(r_{ij})$, $i,j=1,2$, it is convenient to define the $\bar r_{ij}$ to be the restriction to $\mathcal{D}_0=0$ for given $\k$ by setting $\omega$ according to Equation (\ref{disp loci}) for the branch in question. Hence, the $\e_\lambda$ are independent of $\omega$, and therefore
\begin{equation}\label{D_omega}
    \pderiv{D_\lambda}{\omega}=\e_\lambda^\dagger\,\vdot\, \pderiv{\boldR_0}{\omega} \,\vdot\,\e_\lambda=2\omega\, \e_\lambda^\dagger\,\vdot\, \I \,\vdot\,\e_\lambda=2\omega.
\end{equation}
Explicitly, $\e=n_0^{-1}(\bar r_{12},\, -\bar r_{11})^T$, where $n_0=(\bar r_{11}^2+\bar r_{12}^2)^{1/2}$ is the normalizing factor. In terms of the usual variables,
\begin{equation}\label{e}
    \e = \frac{1}{n_0}\left(a^2 k^2 \sin \theta  \cos \theta -c^2 k_x k_z,\, a^2 k^2 \cos ^2\theta +c^2
   k_x^2-\omega ^2\right),
\end{equation}
where the standard MHD dispersion relation (\ref{Dmhd}) can be used to eliminate $\omega$ or $k_z$, as required.

It is now convenient to define the dimensionless \emph{decay rate factor} by scaling $\gamma$ by the wave period:
\begin{equation}\label{Delta}
    \Delta=-\frac{2\pi\,\gamma_\lambda}{\omega} = \frac{\epsilon\,\pi}{\omega^2}\,\e_\lambda^\dagger\,\vdot\,\boldQ(\x,\k_r)\,\vdot\, \e_\lambda >0.
\end{equation}

To understand what this means for local heating, consider a single steady state (driven) wave with energy density $\mathscr{E}(\x)$ (wave energy per unit volume, which depends on the focusing or defocusing of rays). This scales quadratically with local wave amplitude, and so the rate of wave energy decay, and thus the rate of heating (see Appendix \ref{app:elas}), is
\begin{equation}\label{heat}
    \mathscr{H}(\x)=-2\gamma\,\mathscr{E}(\x) = \frac{\omega \Delta}{\pi}\,\mathscr{E}(\x).
\end{equation}
Note that the wave is not decaying \emph{in situ}; the decay is a rate following the ray at the group velocity, but depositing heat locally.\footnote{In the dissipationless ideal (Hermitian) system, the wave action density $\mathscr{I}=\mathscr{E}/\omega$ satisfies the conservation law $\partial\mathscr{I}/\partial t+\Div(\mathscr{I}\,\boldv_g)=0$, where $\boldv_g$ is the group velocity \citep[][Eq.~11.91]{Whi74aa}. This derives from Noether's theorem and system invariance under time translation. Rewriting the conservation equation as $d\mathscr{I}/d t = -\mathscr{I}\,\Div\boldv_g$, where the total time derivative $d/dt=\partial/\partial t+\boldv_g\vdot\grad$ follows along the ray, makes clear the role of convergence and divergence of ray paths in increasing and diminishing $\mathscr{I}$. Incorporating dissipation modifies this to $d\mathscr{I}/d t + \mathscr{I}\,\Div\boldv_g=2\gamma\,\mathscr{I}$, where $\gamma\le0$. In our case, $\omega$ is fixed and the equation determines $\mathscr{E}$ given $\boldv_g$ and the value $\mathscr{E}_0=\mathscr{E}(\x_0)$ at the start of the rays. In the weak dissipation approximation, $\gamma$ is dropped from the conservation equation, but retained in the formula for $\mathscr{H}$. We need not evaluate $\mathscr{E}$ explicitly if the heating rate is expressed in terms of the local energy density, but it must be calculated if an absolute heating rate is required.} 

Although heating is not included in the energy equations (\ref{p c}) and (\ref{p n}), the decay of modes due to collisional terms in the momentum equations (\ref{mmntm c}) and (\ref{mmntm n}) is assumed to ultimately feed back to the atmosphere as heat. Any back-reaction of this onto the modes would be a higher order effect that is in any case neglected in the weak dissipation approximation.

\begin{figure}[htb]
    \centering
    \includegraphics[width=.25\textwidth]{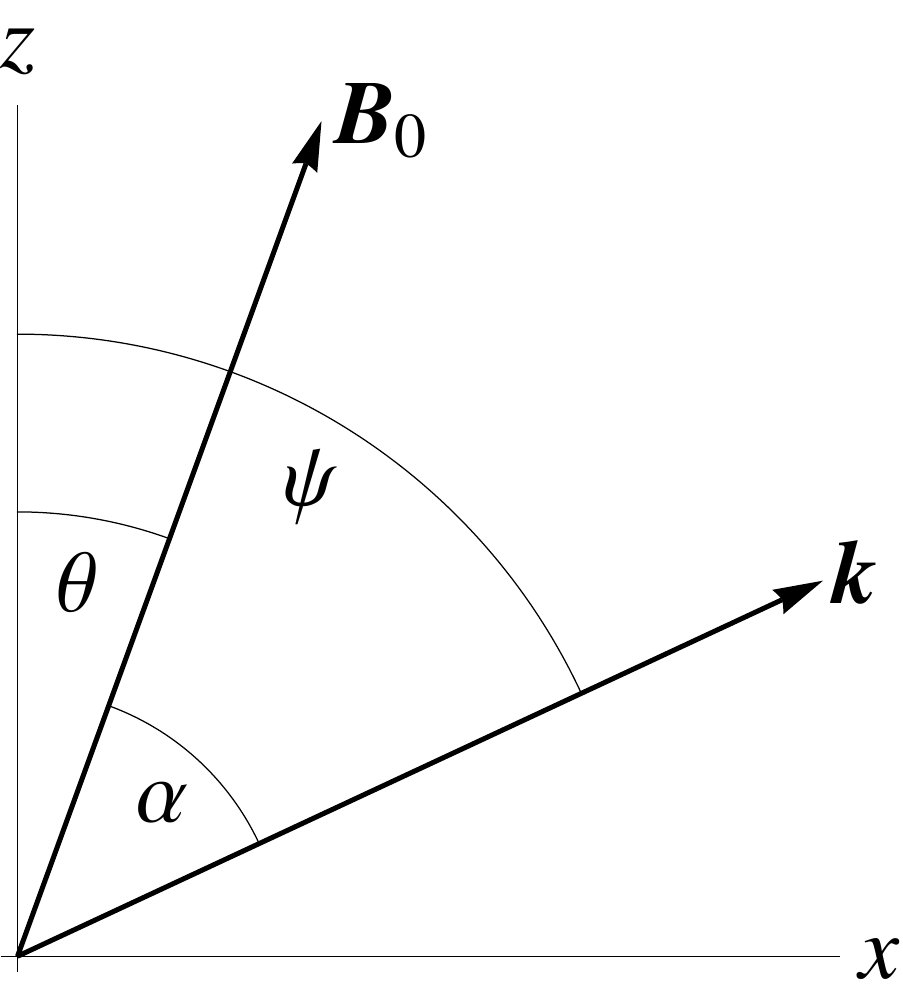}
    \caption{Schematic diagram relating the angles $\theta$, $\psi$ and $\alpha$.}
    \label{fig:angles}
\end{figure}

The dispersion loci in real $(a,k_z)$ space are given implicitly by Equation (\ref{Dmhd}), i.e., 
\begin{equation}\label{disp loci}
    \omega^2 =\half\left(
    (a^2+c^2)k^2\pm k\sqrt{(a^2-c^2)^2k^2+4a^2c^2\kperp^2}\,\,\right),
\end{equation}
where $\kperp=k_x\cos\theta-k_z\sin\theta=k\sin\alpha$, with the positive sign representing the fast wave and the negative sign the slow wave. Again, $\alpha=\psi-\theta$ is the attack angle (Figure \ref{fig:angles}). 

The components of $\boldQ$ are 
\begin{subequations}\label{Q}%
\begin{equation}
    Q_{11} =\frac{ \ac^2 k^2\cos \theta  \left(\omega ^2 \cos \theta -\cn^2 \kperp k_x\right)-2  \cn^4 k^2
   k_x^2+3 \omega ^2 \cn^2 k_x^2-\omega ^4}{(\chi +1) \omega ^2},
\end{equation}
\begin{equation}
    Q_{22} = \frac{ \ac^2 k^2 \sin \theta  \left(\cn^2 \kperp k_z+\omega ^2 \sin \theta \right)+\cn^2 k_z^2
   \left(3 \omega ^2-2  \cn^2 k^2\right)-\omega ^4}{(\chi +1) \omega ^2},
\end{equation}
and
\begin{equation}
    Q_{12} = Q_{21}= \frac{ \ac^2 k^2\sin 2 \theta  \left(\cn^2 k^2 -2 \omega ^2\right)-2 \cn^2 k_x k_z \left(
   \left(\ac^2+4 \cn^2\right)k^2-6 \omega ^2\right)}{4 (\chi +1) \omega ^2}.
\end{equation}
\end{subequations}
It has been convenient to return to the charges-specific Alfv\'en speed $\ac$ and neutrals-specific sound speed $\cn$ here in preference to the total Alfv\'en and sound speeds, which play a more natural role in the MHD (fully coupled) dispersion matrix $\boldR_0$.

\begin{figure}
    \centering
    \includegraphics[width=.95\textwidth]{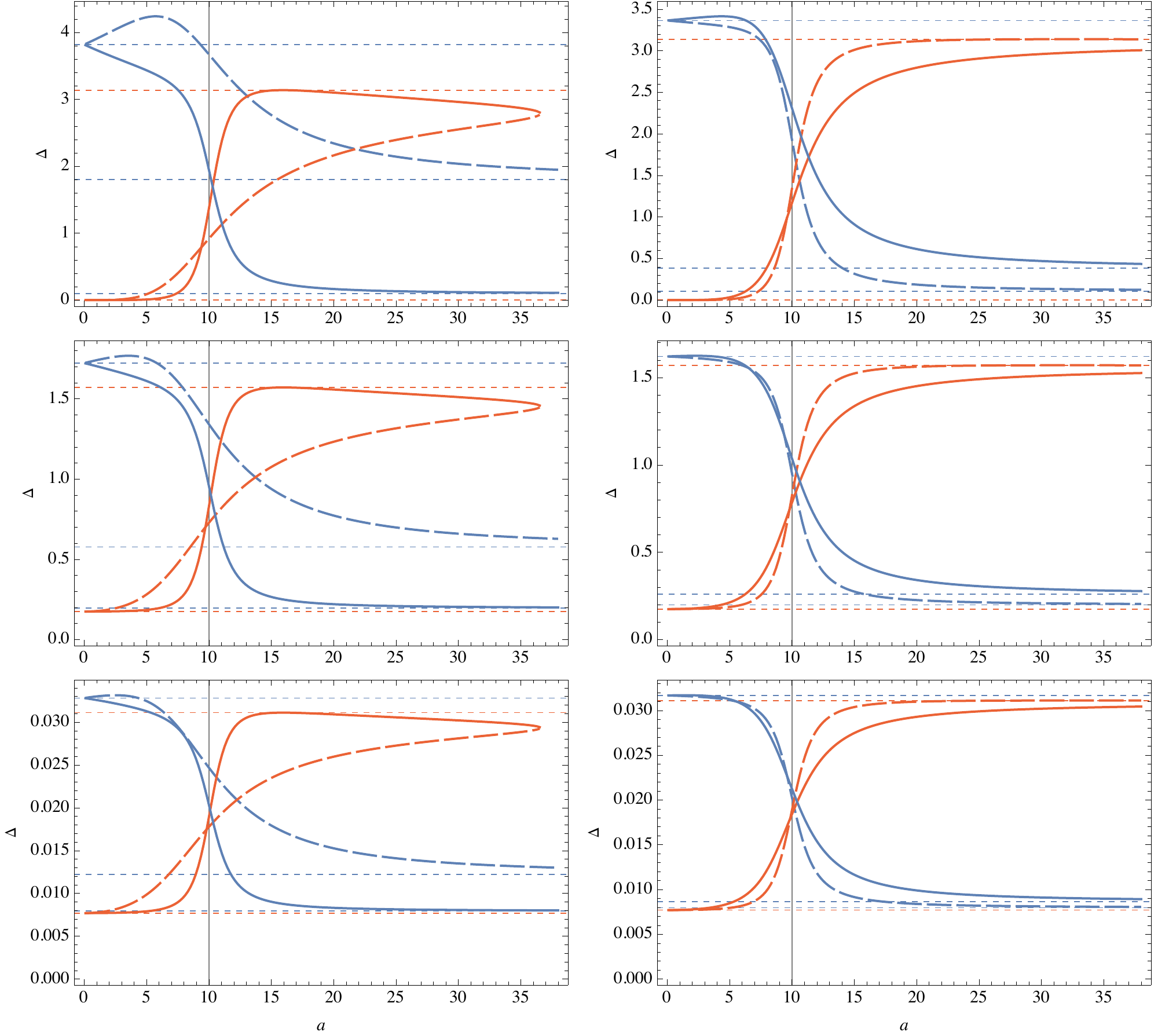}
    \caption{Decay rate factor $\Delta$ as a function of Alfv\'en speed $a$ ($\rm km\,s^{-1}$), for $\epsilon=1$, for the two cases of Figure \ref{fig:star} (left and right panels respectively). The blue curves are for the slow waves and the red curves are for the fast waves. The full curves correspond to the upgoing branches $k_z>0$ and the long-dashed curves to the downgoing branches $k_z<0$. All curves scale linearly with $\epsilon$. The vertical line indicates the position of the $a=c$ equipartition level. Top row: as in the two panels of Figure \ref{fig:star} with $\chi=10^3$. Middle row: same but with $\chi=1$. Bottom row: same but with $\chi=10^{-2}$. The horizontal dashed lines are asymptotic values as defined in Equations (\ref{Delta as}), and are (bottom to top):
    $\Delta_{\text{ac},0}$ (red), $\Delta_{\text{ac},\infty}$ (both, blue), $\Delta_{\text{mag},\infty}$ (red), and $\Delta_{\text{mag},0}$ (blue).}
    \label{fig:Delta}
\end{figure}

The matrix $\boldQ$ is easily evaluated and may be used to calculate $\gamma_\lambda$ by evaluating a simple $2\times2$ quadratic form along each ray. The top row of Figure \ref{fig:Delta} plots the decay rate factor $\Delta$ for the eigenmodes displayed in Figure \ref{fig:star} with $\chi=10^3$ (characteristic of the low chromosphere) and $\epsilon=1$. The full curves represent the upgoing waves, which pass close to the strong conversion region around $a=10$ in the left panel. Not surprisingly, these exhibit a sharp change in $\Delta$ as they pass through conversion and rapidly take on different physical natures, magnetic or acoustic. A slower version of the same process is seen for the downgoing rays, for which the conversion region is weaker and more diffuse. In the right panel it is the dashed curves (negative $k_z$) that change most sharply, as expected since the fast-slow conversion region is most compact for downgoing rays in this case. 

The second and third rows of Figure \ref{fig:Delta} are for the same two cases, but with $\chi=1$ (upper chromosphere) and $\chi=10^{-2}$ (transition region, TR) respectively. Clearly, the effect is much reduced in the TR, across which $\chi$ drops from about 1 to below $10^{-5}$.

Overall, the magnetic wave is damped much more strongly than the predominantly acoustic wave, both when it is slow in $a<c$ and when it is fast in $a>c$, which may also be observed in Figure \ref{fig:kz3D}. This dichotomy is consistent with the expectation from one-fluid MHD with a generalized Ohm's law that ambipolar diffusion (which is a one-fluid ansatz for drift between charges and neutrals) acts predominantly on magnetic terms. However, for moderate $\chi$ (middle row), the acoustic waves still exhibit significant damping.

Of course, the values of $\Delta$ plotted in Figure \ref{fig:Delta} scale linearly with $\epsilon$, so will in practice be much smaller for `low frequency' waves.

Figure \ref{fig:Delta} reveals that the decay rate factors $\Delta$ of the acoustic and magnetic waves each have distinct behaviours in the asymptotic regimes $a\ll c$ and $a\gg c$. These can be found analytically. For $a\gg c$, $\Delta$ for the magnetic wave (fast in this region) approaches\footnote{Of course, this case is not strictly asymptotic, as the fast wave reflects at a finite height, and the $a\to\infty$ limit does not apply. The relevant asymptotic regime is more properly written as $c\ll a\ll a_\text{turn}$, where $a_\text{turn}$ is the Alfv\'en speed at the turning point. With this in mind, $\Delta_{\text{mag},\infty}$ is seen in the figure to be very accurate in the left panels on the close-avoided-crossing branch. The wide crossing branch does not have sufficient room to achieve its asymptotic limit in the cases shown. In the right panels, $a_\text{turn}$ is much larger, about 125, and so is less intrusive.}
\begin{subequations}\label{Delta as}
\begin{equation}\label{Delta mag inf}
    \Delta_{\text{mag},\infty} \sim \frac{\epsilon\,\pi\,\chi}{1+\chi} = \epsilon\,\pi\,\xi_n \quad\text{as $a\to\infty$,}
\end{equation}
where $\xi_n=\rhon/\rho=\chi/(\chi+1)$ is the neutral ionization fraction. This could hardly be simpler. For the acoustic wave (slow)
\begin{equation}\label{Delta ac inf}
    \Delta_{\text{ac},\infty} \sim \frac{\epsilon\,\pi\,\chi}{1+\chi}\,\frac{  \omega^{-2}(\chi +1)^2 \sec^2\theta  \left(c^2 k_x^2 \pm 2 c\,
   k_x\, \omega  \sin \theta +\omega ^2\right)-\chi  (\chi
   +2)}{ (\chi +2)^2} \quad\text{as $a\to\infty$},
\end{equation}
where the `$-$' sign corresponds to the upgoing wave and the `$+$' sign to downgoing.

Conversely, for $a\ll c$, the acoustic (i.e., fast) decay rate factor is asymptotically
\begin{equation}\label{Delta ac 0}
    \Delta_{\text{ac},0} \sim \frac{\epsilon\,\pi\,\chi}{(1+\chi)(2+\chi)^2} = \frac{\epsilon\,\pi\,\xi_n}{(2+\chi)^2} \quad\text{as $a\to0$,}
\end{equation}
and for the magnetic (slow) wave
\begin{equation}\label{Delta mac 0}
      \Delta_{\text{mag},0} \sim \frac{\epsilon\,\pi\,\chi}{1+\chi}\,\frac{ (\chi +1)^2 \sec ^2\theta +2 \chi +3}{
   (\chi +2)^2} \quad\text{as $a\to0$.}
\end{equation}
\end{subequations}
When deriving these results for the slow wave in both asymptotic regimes, we have used $k_z \sim \sqrt{\pm\omega^2/\cT^2 - k_x^2}$ where $\cT=a\,c/\sqrt{a^2+c^2}$ is the cusp speed, which follows from the MHD dispersion relation.

These decay rate asymptotes are indicated by the horizontal dashed lines in Figure \ref{fig:Delta}. They provide very useful analytic estimates of decay rates of all magneto-acoustic wave types in both asymptotic regimes, and in particular show explicitly how they depend on the atmospheric and wave parameters. 

To recover the decay rate in units of $\rm s^{-1}$, just divide $\Delta$ by the wave period, $\gamma=-\omega\,\Delta/2\pi$.

%%%%%%
\subsection{Implications of Mode Conversion for Frictional Decay and Heating}\label{sec:implications}
Mode transmission as given by Equation (\ref{T}) is total at zero attack angle. However, it falls off rapidly away from that direction if $k \, |h_s|=\omega\,|h_s|/c$ is large at the star point. Consequently, transmission of high frequency waves is effectively restricted to a narrow wedge of attack directions, which means that conversion is favoured over transmission at all but this narrow range. This is to be expected, since high frequency implies small wavelength, and hence greater validity of the eikonal approximation, so high-frequency rays on the full dispersion curve are more compactly restricted to it. 

\begin{figure}
    \centering
    \includegraphics[width=.8\textwidth]{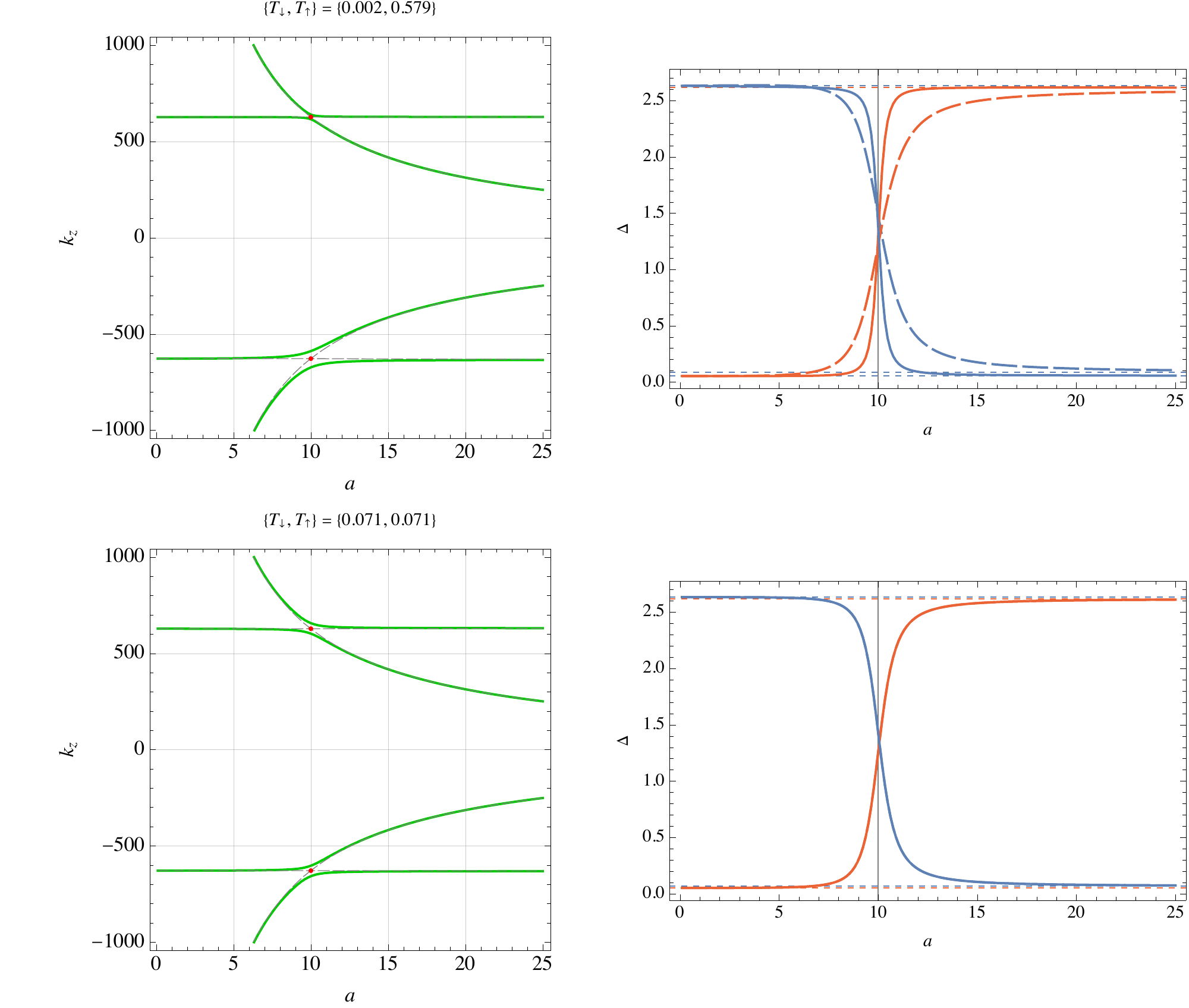}
    \caption{Top row: a high frequency (1 Hz) example with $k_x=30$ $\rm Mm^{-1}$, $c=10$ $\rm km\,s^{-1}$, $\theta=5^\circ$, $\chi=5$, $h=0.15$ Mm, and $\epsilon=1$. Left: the phase space dispersion curves, with downgoing and upgoing transmission coefficients shown at top. Right: decay rate factor $\Delta$ for the same case. For comparison, the transmission coefficients for $\epsilon=0$ are 0.005 and 0.630. Here, magnetic field inclination $\theta$ from the vertical has been chosen to yield strong transmission through the upper star point. By $\theta=7^\circ$, $T_\uparrow$ has already fallen to $0.145$, with $T_\uparrow=1$ at about $\theta=2.75^\circ$. Bottom row: the same but for the 1.5D vertical wave case $k_x=0$, for which conversion is 93\%. The up and down going $\Delta$ curves are coincident. }
    \label{fig:hifreq}
\end{figure}

The consequence of this is that a low-dissipation high frequency acoustic (fast) wave incident on the equipartition layer from below will for most directions convert to a magnetic (still fast) wave on passing through it, and suddenly become subject to the high dissipation that is associated with moderate or large $\epsilon$. This is illustrated in Figure \ref{fig:hifreq}. Note the stark difference between transmission for upgoing and downgoing waves due to their differing attack angles in this instance. Consequently, the upward acoustic wave will be damped very quickly, yielding enhanced local heating beyond where (total) sound and Alfv\'en speeds coincide. 

Except for $\Delta_{\text{ac},\infty}$ the asymptotic expressions (\ref{Delta as}) for decay factor $\Delta$ all scale exactly linearly with frequency, so the heating rate $\mathscr{H}$ scales quadratically with $\omega$ for given wave energy density.

Simulations are often carried out in the 1.5-dimensional (1.5D) vertical wave case, where $k_x=0$ but $\theta\ne0$ or $\pi/2$, so there are $x$-velocities. The second row of Figure \ref{fig:hifreq} illustrates such an example. The $k_z$ curves are up-down symmetric, so the up and down $\Delta$ curves sit one on top of the other. For the high frequency case shown, the field inclination $\theta=5^\circ$ (which in this case is the attack angle) is too large to allow much transmission, so conversion is near-total.

%%%%%%%%%%%%%%%%%%%%%%%%%%%%%%%%%%%%%%%%%%
\section{Conclusions} \label{sec:conc}
The two dimensional 2F plasma supports three (generally) distinct wave types: an acoustic neutral wave and two magneto-acoustic waves, fast and slow. The Alfv\'en wave was ignored in this study by supressing velocities in the ignorable $y$-direction.

The significant results obtained concerning the effects of inter-species collisions on these waves are:
\begin{enumerate}
    \item The acoustic neutral wave, which has the expected dispersion relation $k=\omega/\cn$ in the absence of collisions, instead has 
    \[
      k_z \sim \pm(1+i)\frac{\sqrt{\omega\,\nunc}}{c} \frac{\chi+2}{2\sqrt{\chi+1}}
    \]
    for $\nunc\gg\omega$, resulting in tiny wavelengths and correspondingly tiny decay lengths. The two effects move in lockstep due to the $1+i$ factor. Effectively, this wave is drastically slowed and dissipated over a few short wavelengths in any scenario relevant to the lower solar atmosphere (see Fig.~\ref{fig:nunc}). Only waves with frequencies comparable to the collision frequency can escape this fate. {See Equation (\ref{k3}).}
    \item The effects of collisions on the magneto-acoustic waves are most notably dependent on the dimensionless frequency parameter $\epsilon=\omega/\nunc=\chi\,\omega/\nucn$. ionization ratio $\chi=\rhon/\rhoc$ and wave parameters $\omega$ and $k_x$ play secondary roles, though wave type (magnetically or acoustically dominated) is also crucial.
    \item Mode transmission and conversion near the equipartition level $a=c$ at $z=z_\text{eq}$ is found to behave nearly identically to the 1F MHD case, where transmission is favoured by small attack angle $\alpha$ between the wave vector and the magnetic field. There is a small $\mathcal{O}(\epsilon^2)$ correction due to collisions, but this will be negligible in practical solar circumstances. It may be of relevance in other plasmas where waves close to the collision frequency are more prominent. However, there is an $\mathcal{O}(\epsilon)$ effect on the phase of the converted waves. {See Equations (\ref{tau_beta}), (\ref{eta_star}) and (\ref{T}).}
    \item Parameters of importance to mode conversion include:
    \begin{enumerate}
        \item $\kappa=\pi\,k |h_s|=2\pi^2 |h_s|/\lambda$, which is $2\pi^2$ times the number of wavelengths $\lambda$ that fit across the oblique conversion thickness $h_s$. This must be $\mathcal{O}(1)$ or more for there to be significant conversion $C=1-T$. If $\kappa\gg1$ then conversion is near-total, which is to be expected since a high frequency wave should be more tightly bound to the eikonal dispersion curves (the cyan curves in Figure \ref{fig:star}).
        \item Attack angle $\alpha$. Small $\alpha$ favours transmission at the expense of conversion, which is again to be expected. For example, a pure longitudinal acoustic wave at zero attack angle does not perturb or interact with the magnetic field, and hence should pass through the equipartition level without changing its acoustic identity. As $\alpha$ increases toward $90^\circ$, the maximal conversion allowed by $\kappa$ may take effect.
        \item The \emph{transmission wedge} of attack angles through which there is significant transmission narrows rapidly with increasing frequency, $|\alpha|\lesssim \sqrt{c/\pi\,\omega\,|h_s|}$. At high frequencies, this is so narrow that an only a vanishingly small proportion of an ensemble of random incident acoustic waves would transmit. {See Equation (\ref{T}).}
    \end{enumerate}
    \item Nevertheless, mode conversion plays a very important role in the overall story. It is found that magnetically dominated waves (i.e., slow waves where $a<c$ and fast waves where $a>c$) are much more susceptible to collisional damping than acoustically dominated magneto-acoustic waves. Most strikingly, the local heating rate per unit volume for magnetic waves on $a\gg c$ is simply $\epsilon\,\omega\,\xi_n\,\mathscr{E}$, where $\mathscr{E}$ is the local wave energy density. For acoustic waves launched from below the equipartition level, this means that the converted (acoustic-to-magnetic) portion of the incident waves will become subject to enhanced collisional damping on reaching and passing through $z_\text{eq}$. But we saw in the previous point that nearly all high frequency acoustic waves incident on $z_\text{eq}$ from below convert, so nearly the entire wave flux is susceptible to enhanced dissipation. {See Equations (\ref{Delta as}).}
    \begin{enumerate}
        \item As simple as the asymptotic decay parameters listed in Equations (\ref{Delta as}) are, the full exact formula for $\Delta$ is hardly more complicated. It simply involves finding the null vector of a $2\times2$ singular matrix, and multiplying out a $2\times2$ quadratic form. {See Equation (\ref{Delta}).}
    \end{enumerate}
    
\end{enumerate}

Let us now make a crude and speculative accounting of the energy budget of the upper chromosphere. \citet{WitNoy77aa} estimate quiet Sun upper chromospheric radiative losses at 300 $\rm W\, m^{-2}$. Adopting atmospheric model C7 of \citet{AvrLoe08aa}, and supposing that the tabulated turbulent velocities $V_\text{turb}(z)$ can be represented as or behave similarly to high frequency waves, we have the wave energy density $\mathscr{E}\sim\rho\,V_\text{turb}^2$. Using Equation (\ref{Delta mag inf}) for $\Delta$, assuming we are above the $a=c$ canopy, the collisional heating rate of Equation (\ref{heat}) is then $\mathscr{H}(z)\sim4\pi^2\,\nu^2\,\xi_n(z)\,\mathscr{E}(z)/\nunc(z)$, where $\nu=\omega/2\pi$ is the wave frequency in Hz. If we integrate this over $1000\,{\rm km}<z<2000\,{\rm km}$ using the $\nunc$ of Figure \ref{fig:nunc}, we get a total heating of $314\,\nu^2\,\rm W\,m^{-2}$, which for 1 Hz or above is comparable to the required losses. In more generality, a turbulent distribution with significant energy at around 1 Hz would be a potential source of heating. Similar conclusions are drawn for collision-damped Alfv\'en waves by \citet{De-MarHud01aa} and \citet{SonVas11aa}. Of course, there are many crude assumptions in this line of reasoning, but it does at least suggest that there may be a sufficient store of energy and a viable dissipation mechanism to make a significant contribution to supplying radiative losses.

On the other hand, there is also a ubiquitous bath of low frequency (2--10 mHz) compressible and incompressible wave energy in the solar chromosphere  \citep{ZaqErd09aa,MorVerJes12aa,JesMorVer15aa}, much of it in flux tube structures, that the turbulent velocity estimates of classical empirical models such as C7 do not capture. This energy is inaccessible directly to the 2F collisional or 1F ambipolar diffusion dissipation processes due to its very large length scales. However, even such long-period waves can drive turbulence \citep{MatZanOug99aa} that displays much smaller observationally-inaccessible scales, and which may be subject to these mechanisms. This turbulence may be episodic and exist in addition to the basal values invoked in steady state atmospheric models. Smaller scales are also present in shocks and other dynamic events not treated here that may themselves drive turbulence \citep{ReaLepCar08aa}.

The sound speed in the upper solar chromosphere is about 10 $\rm km\,s^{-1}$ which translates to a wavelength of about 10 km for 1 Hz acoustic waves. Assuming a moderate magnetic field strength of $10^{-3}$~T (10 G), the Alfv\'en speed at the top of the chromosphere is about 80 $\rm km\,s^{-1}$ with a roughly 80 km wavelength at 1 Hz for magnetically dominated fast waves. Both of these length scales are accessible computationally in simulations, especially the fast wave which is most susceptible to collisional damping. Turbulence may be expected to have even shorter length and time scales and therefore potentially be more easily damped. In future work we will test the theoretical conclusions derived here using 2F simulations.

%%%%%%%%%%%%%%%%%%%%%%%%%%%%%%%%%%%%%%%%%%%%%%
\appendix
\section{Dispersion Matrix Components}\label{app:R}
Explicitly, the components of $\mathbf{R}$ are
\begin{subequations}\label{R}
\begin{multline}\label{R11}
    R_{1\,1}=\omega ^2-\left(a^2 \cos ^2\theta +c^2\right)k_x^2 
    -a^2 k_z^2 \cos ^2\theta 
   \\
   \quad -i\, \epsilon 
   \left(\frac{a^2 c^2 k^2 k_x^2 (\chi +1) \cos ^2\theta 
   }{(\chi +2) \omega ^2}-\frac{a^2 c^2 k^2 k_x k_z (\chi +1)
   \sin 2\theta    }{2(\chi +2) \omega
   ^2}-a^2 k^2 \cos ^2\theta +\frac{2 c^4 k^2 k_x^2(\chi +1)
   }{(\chi +2)^2 \omega ^2}-\frac{3 c^2 k_x^2}{\chi
   +2}+\frac{\omega ^2}{\chi +1}\right);
\end{multline}
\begin{multline}\label{R22}
    R_{2\,2}=\omega ^2-a^2 k^2 \sin ^2\theta
   -c^2 k_z^2\\
   \quad -i\, \epsilon  \left(\frac{a^2 c^2
   k^2k_z^2 (\chi +1) \sin ^2\theta  }{(\chi +2) \omega
   ^2}-\frac{a^2 c^2 k^2 k_x k_z(\chi +1) \sin 2\theta 
    }{2(\chi +2) \omega ^2}-a^2 k^2 \sin ^2\theta +\frac{2 c^4 k^2k_z^2 (\chi +1)
   }{(\chi +2)^2 \omega ^2}-\frac{3 c^2 k_z^2}{\chi
   +2}+\frac{\omega ^2}{\chi +1}\right);
\end{multline}
%\newpage
\begin{multline}\label{R12}
    R_{1\,2}=a^2 k^2 \sin \theta  \cos \theta -c^2
   k_x k_z\\
   -\frac{i\, \epsilon}{(\chi +2)^2 \omega
   ^2}  \left(a^2 c^2 k^2 k_x k_z(\chi +1) (\chi
   +2) \sin ^2\theta -\half a^2 c^2 k^2 k_x^2(\chi +1) (\chi +2) \sin
   2\theta     +\half\omega ^2 a^2 k^2 (\chi +2)^2 
   \sin 2\theta   \right.\\
  \left. +2 c^4 k^2 k_x k_z(\chi +1) -3
   \omega ^2 c^2 k_x k_z (\chi +2)  \right);
\end{multline}
and
\begin{multline}\label{R21}
    R_{2\,1}=a^2 k^2 \sin \theta  \cos \theta -c^2
   k_x k_z\\
   -\frac{i\, \epsilon}{(\chi +2)^2 \omega ^2}  \left( a^2 c^2 k^2  k_x k_z (\chi +1) (\chi +2) \cos ^2\theta-\half a^2 c^2 k^2 k_z^2(\chi +1) (\chi +2) \sin
   2\theta  + \half\omega ^2 a^2 k^2 (\chi +2)^2  \sin 2\theta   \right.\\
   \left.
   +2 c^4 k^2 k_x
   k_z (\chi +1) -3\omega ^2 c^2 k_x k_z(\chi +2)  \right),
\end{multline}
\end{subequations}
where $\theta$ is the angle of the magnetic field from the vertical.

%%%%%%%

\section{Energy dissipation in linear elastic collisions}\label{app:elas}
Braginskii's elastic relations \citep{Bra65aa} ensures the `elasticity' of the collisions by imposing a specific form to the collisional terms that conserves energy and momentum. And yet the dispersion relation Equation (\ref{D full}) has complex roots corresponding to decay, which leads to the counter-intuitive conclusion that linear elastic collisions are actually inelastic. This can be explained by extending the wave energy constructions of \citet{Eck63aa} and \citet{BraLou74aa} to two fluids.

Beginning with the 2F linearized continuity, momentum, energy and induction equations as in Equations (\ref{basiceqns}), and retaining gravity $g$ in the $-z$ direction for completeness, a quadratic wave energy equation may be constructed by algebraic manipulation
\begin{gather}
    \frac{\partial}{\partial t}(W_\text{n} + W_\text{c}) + \Div (\mathbf{s}_\text{n} + \mathbf{s}_\text{c}) = - \alpha_\text{cn} |\mathbf{v}_\text{c} - \mathbf{v}_\text{n}|^2,\label{Eckart E}\\[4pt]
\intertext{where $W_\text{n}$ and $W_\text{c}$ are the energy densities of the two species and $\bolds_\text{n}$ and $\bolds_\text{c}$ are the corresponding wave energy flux densities (energy per unit area per unit time), with}
    W_\text{n} = \frac{1}{2}\rho_\text{n}v_n^2 + \frac{p_\text{n}^2}{2\rho_\text{n}c_\text{n}^2} - \left(\frac{g^2\rho_\text{n}}{2c_\text{n}^2} + \frac{g}{2}\frac{d \rho_\text{n}}{dz} \right)\xi_{\text{n},z}^2,\\[4pt]
    W_\text{c} = \frac{1}{2}\rho_\text{c}v_c^2 + \frac{p_\text{c}^2}{2\rho_\text{c}c_\text{c}^2} - \left(\frac{g^2\rho_\text{c}}{2c_\text{c}^2} + \frac{g}{2}\frac{d \rho_\text{c}}{dz} \right)\xi_{\text{c},z}^2 + \frac{b^2}{2\mu_0},\\[4pt]
    \mathbf{s}_\text{n} = p_\text{n} \mathbf{v}_\text{n},\\[4pt]
    \mathbf{s}_\text{c} =  p_\text{c} \mathbf{v}_\text{c} + \frac{1}{\mu_0}\left[(\mathbf{B}\, \vdot\, \mathbf{b}) \mathbf{v}_\text{c} - (\mathbf{b}\, \vdot\, \mathbf{v}_\text{c}) \mathbf{B}\right].   
\end{gather}
Here $\xi_{\alpha,z}$ is the $z$-component of the plasma displacement vector $\bxi_\alpha$ of species $\alpha$. The terms in the energies are respectively the kinetic, compressional, buoyancy and (for the charges) magnetic energy densities. The terms in the fluxes are the rate of working of the gas pressure perturbations and (for the charges) the Poynting flux. Note that collisional terms were included in the momentum equations only, disregarding those in the energy and induction equations.

From the more general 2F equations including all collisional contributions, two-fluid collisional heating appears as a source term in the internal energy equation, and is given by \citep[Eqs.~(25)]{PopLukKho19aa}:
\begin{gather}
    Q_\text{n} = \frac{1}{2}\tilde\alpha_\text{cn} |\mathbf{v}_\text{c} - \mathbf{v}_\text{n}|^2 + \frac{1}{\gamma_\text{a} - 1}\frac{k_\text{B}}{m_\text{n}}\tilde\alpha_\text{cn}(T_\text{c} - T_\text{n}),\\
    Q_\text{c} = \frac{1}{2}\tilde\alpha_\text{cn} |\mathbf{v}_\text{c} - \mathbf{v}_\text{n}|^2 - \frac{1}{\gamma_\text{a} - 1}\frac{k_\text{B}}{m_\text{n}}\tilde\alpha_\text{cn}(T_\text{c} - T_\text{n}).
\end{gather}
The first term in the equations is called the frictional heating (FH) and the second one the thermal exchange (TE), which simply shifts thermal energy between species. Then, the total collisional heating is obtained as the sum 
\begin{equation*}
    Q_\text{n} + Q_\text{c} = \tilde\alpha_\text{cn} |\mathbf{v}_\text{c} - \mathbf{v}_\text{n}|^2,
\end{equation*}
where $\tilde\alpha_\text{cn}$ is the nonlinear collisional frequency that reduces to $\alpha_\text{cn}$ in the linear regime.
This then exactly balances the right hand side wave energy sink term in Equation (\ref{Eckart E}).

The fact that this total heating corresponds to the wave energy loss term constructed from equations of the form (\ref{basiceqns}) with collisional terms only in the momentum equations demonstrates that the wave energy decay rate may be interpreted as heating, as in Equation (\ref{heat}). Overall, energy is conserved provided thermal energy is included in the accounting. It is not conserved in the wave energy alone, despite the collisions being elastic.

%%%%%%%%%%%%%%%%
\section{Weak Dissipation of Rays}\label{app:weakz}
This method applied in this appendix is adapted from Section 3.5.1 of \cite{TraBriRic14aa}, but with a twist, some different notation, and a little more explanation of intermediate steps. The aim is to determine the decay rate of a ray due to collisional effects, beyond the geometric focusing and defocusing of real rays. To do so, we regard $\epsilon$ as a small perturbation to the ideal real-ray case.

Let $\boldR=\RH+\RA$ represent the separation of the full dispersion matrix $\boldR(\x,\k)$ into Hermitian and skew-Hermitian parts as before. It transpires that $\RA=i\,\epsilon\,\boldQ$ for real symmetric matrix $\boldQ$ with components set out in Equations (\ref{Q}). Frequency $\omega$ is assumed fixed and so not mentioned explicitly in the arguments.

Also, as seen previously (see Equation (\ref{RH})), $\RH=\boldR_0+i\,\epsilon\,E\,\J$ where $\boldR_0$ is the MHD dispersion matrix. Although it is not strictly necessary to split the $\epsilon\,E$ term from $\boldR_0$, as $\RH$ itself is Hermitian, it is convenient and consistent to do so to separate the $\mathcal{O}(1)$ and $\mathcal{O}(\epsilon)$ terms (this is the twist mentioned above).

The real rays $(\x(t),\k_r(t))$ then derive from $\boldR_0$ alone, and are as in MHD. Let $\k=\k_r+i\,\k_i$ be the \emph{complex} wave vectors determined by the full dispersion relation $\mathcal{D}(\x,\k)=0$. It is to be expected that $\k_i=\mathcal{O}(\epsilon)$, so we let $\k_i=\epsilon\,\bkappa$ where $\bkappa$ is real and of order 1. Then $\boldR(\x,\k)=\boldR_0(\x,\k_r+i\,\epsilon\,\bkappa)+i\,\epsilon\,E(\x,\k_r+i\,\epsilon\,\bkappa)\,\J+ i\,\epsilon\,\boldQ(\x,\k_r+i\,\epsilon\,\bkappa)$. Assume that point $(\x,\k_r)$ is on a real ray with non-degenerate unit (column) eigenvector $\e(\x,\k_r)$, which therefore spans the null space of $\boldR_0(\x,\k_r)$. Projecting the eigenvector onto $\boldR(\x,\k_r)$ from both sides, adopting vector/dyadic notation,
\begin{equation}\label{projR}
    \e^\dagger(\x,\k_r)\,\vdot \bigl[\boldR_0(\x,\k_r+i\,\epsilon\,\bkappa) +i\,\epsilon\, E(\x,\k_r+i\,\epsilon\,\bkappa)\,\J+ i\,\epsilon\,\boldQ(\x,\k_r+i\,\epsilon\,\bkappa)\bigr]\vdot\,\e(\x,\k_r)=0.
\end{equation}
Linearizing in $\epsilon$, and noting that the zeroth order term vanishes since $\e$ is in the null space of $\boldR_0$ leaves
\begin{equation}\label{2proj}
\e^\dagger\vdot\bigl[(\bkappa\vdot\,\partial_{\k_r})\boldR_0(\x,\k_r)+E(\x,\k_r)\,\J+\boldQ(\x,\k_r)\bigr]\vdot\, \e=0.
\end{equation}
Since the rays are real, so will be their eigenvectors $\e$. But $\J$ is real anti-symmetric, so $\e^\dagger\,\vdot\,\J\,\vdot\,\e=0$. Hence the term $E$ plays no dissipative role. This is to be expected, since it forms part of the Hermitian matrix $\RH$.

Labelling our null vectors $\e_\lambda$ by Greek subscripts, $\lambda=\alpha,\,\beta,\, \ldots$ say, and their corresponding disjoint dispersion functions (eigenvectors) by $D_\lambda$, we have $\e_\lambda^\dagger\,\vdot\,\RH\,\vdot\,\e_\lambda=D_\lambda$. Given that $\RH\,\vdot\,\e_\lambda=0$ and $\e_\lambda^\dagger\vdot\,\RH=0$ on the ray by construction, the eigenvectors may be brought inside the $\k$-derivative to yield
\begin{equation}\label{2Dproj}
    (\bkappa\,\vdot\,\partial_{\k_r}) D_\lambda + \e_\lambda^\dagger\vdot\,\boldQ(\x,\k_r)\,\vdot\, \e_\lambda = 0.
\end{equation}

The classical ray equation
\begin{equation}\label{ray_x}
    \deriv{\x}{t} = -\frac{\partial D/\partial\k}{\partial D/\partial\omega}, %\qquad
    %\deriv{\k}{t} = \frac{\partial D/\partial\x}{\partial D/\partial\omega}
\end{equation}
\citep[Eq.~(149)]{Wei62aa} may then be applied to $D_\lambda$ to reduce Equation (\ref{2Dproj}) to
\begin{equation}
   \k_i\,\vdot\,\deriv{\x}{t} = \epsilon\,\frac{\e_\lambda^\dagger\,\vdot\,\boldQ(\x,\k_r)\,\vdot\, \e_\lambda}{\partial D_\lambda/\partial\omega} = -\gamma_\lambda\left(\x(t),\k_r(t)\right)
\end{equation}
along the ray $\lambda$, where we have returned to the original $\k_i$ and defined $\gamma_\lambda$. 

%%%%%%%%%%%%%%%%%%%%%%%%%%%%%%
\begin{acknowledgements}
    This work was supported by the European Research Council through the Consolidator Grant ERC-2017-CoG-771310-PI2FA and by the Spanish Ministry of Science through the grant PID2021-127487NB-I00. M.M.G.M. acknowledges support from the Spanish Ministry of Science and Innovation through the grant CEX2019-0000920-S-20-1 of the Severo Ochoa Program and from the School of Mathematical Sciences at Monash University for sponsoring the visa for his visit during which this work was carried out. We also acknowledge all the suggestions from Elena Khomenko and David Mart\'inez-G\'omez, who shared with us their up-to-date insights on the physical background of the Sun.
\end{acknowledgements}

%%%%%%%%%%%%%%%%%%%%%%%%%%%%%%%%%%%%%%%%%%%%%%%%%%%%%%%%%%%%%%%%%%%%%%%%%%%%%%%%%%%%%%%%%%%%%
%%%%%%%%%%%%%%%%%%%%%%%%%%%%%%%%%%%%%%%%%%%%%%%%%%%%%%%%%%%%%%%%%%%%%%%%%%%%%%%%%%%%%%%%%%%%%
%%%%%%%%%%%%%%%%%%%%%%%%%%%%%%%%%%%%%%%%%%%%%%%%%%%%%%%%%%%%%%%%%%%%%%%%%%%%%%%%%%%%%%%%%%%%%
%%%%%%%%%%%%%%%%%%%%%%%%%%%%%%%%%%%%%%%%%%%%%%%%%%%%%%%%%%%%%%%%%%%%%%%%%%%%%%%%%%%%%%%%%%%%%


\bibliography{fred}
\bibliographystyle{aasjournal}



\end{document}